\documentclass[a4paper]{article}

\usepackage{verbatim}
\usepackage{amsmath,amssymb}
\usepackage{enumerate}
\usepackage{color}

\newtheorem{defn}{Definition}%[section]
\newtheorem{lem}{Lemma}
\newtheorem{prop}{Proposition}
\newtheorem{propdef}[defn]{Proposition-definition}
\newtheorem{thm}{Theorem}
\newtheorem{cor}{Corollary}

\newtheorem{rem}{Remark}

\newenvironment{dem}{{\bf Proof:}}{ { \hspace {\stretch{1} }$\Box$}}

\renewcommand{\d}{\mathrm{d}}
\newcommand{\supp}{\mathrm{supp}}
\newcommand{\Hess}{\mathrm{Hess}}
\renewcommand{\div}{\mathrm{div}}

\newcommand{\N}{{\mathbb N}}
\newcommand{\Z}{{\mathbb Z}}

\newcommand{\R}{{\mathbb R}}
\newcommand{\C}{{\mathbb C}}

\newcommand{\Bb}{{\mathcal B}}
\newcommand{\Cc}{{\mathcal C}}

\newcommand{\Hh}{{\mathcal H}}
\newcommand{\Ii}{{\mathcal I}}

\newcommand{\Nn}{{\mathcal N}}
\newcommand{\Ss}{{\mathcal S}}

\newcommand{\W}{{\mathcal W}}
\newcommand{\eps}{\varepsilon}
\renewcommand{\leq}{\leqslant}
\renewcommand{\geq}{\geqslant}

\newcommand{\h}{\eps}

\begin{document}

\title{\bf Global $L^2$-Boundedness Theorems for Semiclassical Fourier Integral Operators with Complex Phase}
\author{Vidian ROUSSE and Torben SWART \\
{\it \small Freie Universit\"at Berlin }}

\date{ }

\maketitle

{\small{\bf Abstract}: In this work, a class of semiclassical Fourier Integral Operators (FIOs) with complex phase associated to some canonical transformation of the phase space $T^*\R^d$ is constructed. Upon some general boundedness assumptions on the symbol and the canonical transformation, their continuity (as operators) from the Schwartz class into itself and from $L^2$ into itself are proven.
}

%%%%%%%%%%%%%%%%%%%%%%%%%%%%%%%%%%%%%%%%%%%%%%%%%%%%%%%%%%
\section{Introduction}
%%%%%%%%%%%%%%%%%%%%%%%%%%%%%%%%%%%%%%%%%%%%%%%%%%%%%%%%%%

We consider semiclassical ($\h\in(0,1]$ will be the small parameter) oscillatory integral operators on $\R^d$ formally given by
\begin{equation} \label{eq:def}
\varphi\mapsto\left[\Ii^\h(\Phi;u)\varphi\right](x):=\frac{1}{(2\pi\h)^{(d+D)/2}}\int_{\R^d}\int_{\R^D}e^{\frac{i}{\h}\Phi(x,y,\eta)}u(x,y,\eta)\varphi(y)\d\eta\d y
\end{equation}
where
\begin{itemize}
\item
$\Phi$ is a smooth real or complex-valued phase function,
\item
$u$ is a smooth complex-valued amplitude belonging to some symbol class.
\end{itemize}
An abundant literature is now available about numerous properties of such operators with different assumptions on the phase $\Phi$ and the amplitude $u$, we will mainly quote articles related to our work sometimes disregarding other interesting properties. The general formulation \eqref{eq:def} includes as typical examples two families of operators with $D=d$ that have been intensively studied since the '70s.

The first family corresponds to the choice $\Phi(x,y,\eta)=(x-y)\cdot\eta$ and consists of semiclassical Pseudo-Differential Operators (PDOs) (see for instance \cite{[Martinez]}). Those turned to be very relevant to produce parametrices of Partial Differential Equations (PDEs), for instance of elliptic type, with the help of symbolic and functional calculus. As far as global results are concerned, Calder\'on and Vaillancourt \cite{[CalderonVaillancourt]} showed a fundamental result: PDOs are bounded operators from $L^2(\R^d;\C)$ into itself if the amplitude and its derivatives up to some order are globally bounded.

The second family corresponds to the choice $\Phi(x,y,\eta)=S(x,\eta)-y\cdot\eta$ where the real-valued function $S$ is a generating function of some canonical transformation $\kappa$ of the phase space $T^*\R^d$ {\it i.e.}
\begin{equation*}
\kappa(\nabla_\eta S(x,\eta),\eta)=(x,\nabla_x S(x,\eta)).
\end{equation*}
They are the so-called semiclassical Fourier Integral Operators (FIOs) with real-valued phase associated to a canonical tranformation (see for instance \cite{[Robert]}) and are widely used to produce parametrices and semiclassical approximations for evolution equations like wave or Schr\"odinger equations. In the case where $\kappa$ is the identity, we can choose $S(x,\eta)=x\cdot\eta$ and recover the PDOs. The local existence of a generating function $S$ for $\kappa(y,\eta)=(X^\kappa(y,\eta),\Xi^\kappa(y,\eta))$ around a point $(y_0,\eta_0)=\kappa^{-1}(x_0,\xi_0)$ completely relies on the invertibility of the matrix $(\partial_{y_j}X^\kappa_k)_{jk}$ near $(y_0,\eta_0)$. When it is not the case, we usually call $(y_0,\eta_0)$ a turning point of $\kappa$ in the position representation (this is an expression of the problem of ``caustics'' originally exhibited in geometrical optics). To give an idea of how severe this problem can be, we simply mention that even in the case of linear canonical transformations the entire phase space might be constituted of turning points (think about $\kappa(y,\eta)=(\eta,-y)$). As implicitly suggested and first performed by Maslov \cite{[Maslov]}, a local change of representation (for example to the momentum representation) allows to circumvent this difficulty. However, this procedure leads to FIOs whose representation by a single integral is often only local. The extension of the Calder\'on-Vaillancourt result to those FIOs with real-valued phase can be found in \cite{[Robert]}.

The study of even more general FIOs of the form \eqref{eq:def} goes back to the pioneer works \cite{[Hormander]} and \cite{[DuistermaatHormander]}. As for $L^2$-boundedness results, local results (compactly supported amplitudes) were first proven in \cite{[Hormander]} whereas different type of global results depending on the assumptions on $\Phi$ can be found in \cite{[Hormander1]} or \cite{[Fujiwara]} for $D=0$, \cite{[KumanoGo]}, \cite{[Coriasco]} or \cite{[RuzhanskySugimoto]} for $D=d$ and finally \cite{[AsadaFujiwara]} for general $D$.

The aim of this article is essentially to show that for some semiclassical FIOs with complex-valued phase associated to a canonical tranformation the $L^2$-boundedness property still holds with $\h$-independent norm bound. More precisely, we consider $D=2d$, set $\eta=(q,p)$ and make typically the following assumptions on phases and amplitudes:
\begin{itemize}
\item
$\Phi$ is a complex-valued phase function of the form
\begin{eqnarray*}
\Phi^\kappa(x,y,q,p) & = & S^\kappa(q,p)+\Xi^\kappa(q,p)\cdot(x-X^\kappa(q,p))-p\cdot(y-q) \\
 & & +\frac{i}{2}|x-X^\kappa(q,p)|^2+\frac{i}{2}|y-q|^2
\end{eqnarray*}
where $(X^\kappa(q,p),\Xi^\kappa(q,p))$ is the decomposition in position and momentum of the canonical transformation $\kappa(q,p)$ and $S^\kappa(q,p)$ is some real-valued function reminiscent of the action of classical dynamics (see Definition~\ref{def:action} for a precise definition and some properties),
\item
$u$ is a smooth complex-valued function in the symbol class $S[0;4d]$ {\it i.e.}
\begin{equation*}
\forall\alpha\in\N^{4d}, \qquad \Vert\partial^\alpha_{(x,y,q,p)}u\Vert_{L^\infty}=\sup_{(x,y,q,p)\in\R^{4d}}|\partial^\alpha_{(x,y,q,p)}u(x,y,q,p)|<\infty,
\end{equation*}
\item
$\kappa$ is a canonical transformation of class $\Bb$ {\it i.e.} the Jacobian matrix $F^\kappa$ is in $S[0;2d]$ (see Definitions~\ref{def:CanonicalTransform} and~\ref{def:classB} for more precision).
\end{itemize}
The main result now reads.

{\bf Theorem} {\it
Upon preceding assumptions, the operator $\Ii^\h(\Phi^\kappa;u)$ defined by~\eqref{eq:def} is continuous from $\Ss(\R^d;\C)$ into itself and can be uniquely extended to a bounded operator on $L^2(\R^d;\C)$ such that
\begin{equation*}
\Vert \Ii^\h(\Phi^\kappa;u)\Vert_{L^2\to L^2}\leq C(\kappa)\sum_{|\alpha|\leq 4d+1}\Vert\partial^\alpha_{(x,y)}u\Vert_{L^\infty}.
\end{equation*}
}

Recent semiclassical contributions closely related to our work are \cite{[LaptevSigal]}, \cite{[BilyRobert]} and \cite{[Butler]} where the authors provide uniform approximations to the solution of the time-dependent Schr\"odinger equation but all of them considered symbols compactly supported in some of the variables. One of the major improvement of our construction and proof is the removal of this restrictive hypothesis which enforced them to consider unitary propagators only in a truncated way.

More general considerations about non-semiclassical FIOs with complex phase are given in \cite{[MelinSjostrand]}, \cite{[MelinSjostrand1]} and \cite{[Hormander2]} for local properties, \cite{[CordobaFefferman]}, \cite{[Unterberger]}-\cite{[Unterberger1]}, \cite{[Bony]}-\cite{[Bony1]} and \cite{[Tataru]} for global properties.

To summarize, the approach presented here combines two different global advantages: the amplitude needs not be compactly supported in momentum and the problem of turning points does not show up in this setting. Moreover, we get explicit control in $\eps$ for the operator norm and the connection with the underlying geometry of phase space is rather transparent.

An application of those FIOs to approximated propagators of Schr\"odinger equations with subquadratic potential will be provided in the forthcoming article \cite{[SwartRousse]} in the spirit of \cite{[LaptevSigal]}, \cite{[BilyRobert]} or \cite{[Butler]}. To the authors' knowledge, this would be the first mathematical proof of a widely spread and used formal result from Theoretical Chemistry originally named after Herman and Kluk.

The plan of this article is as follows. In Section~\ref{s_symbol}, we recall the definitions and main properties of symbol classes, we introduce the notion of action associated to a canonical transformation which will almost play the role devoted to generating function in the theory of FIOs with real phase, we remind the notion of diffeomorphism of class $\Bb$ as introduced in \cite{[Fujiwara1]} and restrict it to the case of canonical transformation. In Section~\ref{s_Wick}, we present a first construction of FIOs associated to a canonical transformation based on the FBI transform and relate it to Anti-Wick quantization. In Section~\ref{s_FIO}, we introduce a more general notion of FIOs associated to a canonical transformation, show that they are continuous from the class of Schwartz functions into itself (see Theorem~\ref{C0S}) and finally prove the main result of the article, Theorem~\ref{theo:L2bound} which states the $L^2$-boundedness for bounded symbols. In terms of technicalities, the proof of the Schwartz continuity is comparable with the corresponding result for pseudodifferential operators as presented in \cite{[DimassiSjostrand]} or \cite{[Martinez]} whereas the $L^2$-boundedness result uses the strategy of \cite{[Fujiwara]}, \cite{[AsadaFujiwara]} and \cite{[Hormander1]}.

We close this introduction by a short discussion of the notation we use. Throughout this paper, we will use column vectors. We will denote the inner product of two vectors $a, b\in\R^D$ as $a\cdot b:=\sum^D_{j=1}a_jb_j$ which we extend to vectors of $\C^D$ by the same formula. The Hilbert norm of $\C^D$ will be denoted by $|a|:=(\overline{a}\cdot a)^{1/2}$. The transpose of a (real or complex) square matrix $A$ will be $A^\dagger$, whereas $A^*:=\bar{A}^\dagger$ and $I$ stands for the identity matrix. When dealing with diffeomorphisms of $T^*\R^d$ or operators on $L^2(\R^d;\C)$, ${\rm Id}$ stands for the identity morphism or operator. We will use the standard multi-index notation. Following~\cite{[Butler]} and~\cite{[LaptevSigal]}, we will sometimes use subscript to denote differentiation. Thus, for a differentiable mapping $F\in C^1(\R^D,\C^D)$, $F_x(x)$ will denote the transpose of its Jacobian, i.e. $(F_x(x))_{jk}=\partial_{x_j}F_k(x)$. As a crucial example, the factor $X^\kappa_q$ in~\eqref{eq:action} stands for the matrix $(\partial_{q_j}X^\kappa_k)_{1\leq j,k\leq d}$ so that we have the following identity of column vectors $(X.\Xi)_q=X_q\Xi+\Xi_qX$. The Hessian matrix of a mapping $F\in C^2(\R^D,\C)$ will be denoted by $\textrm{Hess}_{x}F(x)$ and the divergence of a mapping $F\in C^1(\R^D,\C^D)$ by $\div_x F$.

%%%%%%%%%%%%%%%%%%%%%%%%%%%%%%%%%%%%%%%%%%%%%%%%%%%%%%%%%%%%%%%%%%%%%%%%%%%%%%%%%%%%%%%%%
\subsection*{Acknowledgement}
%%%%%%%%%%%%%%%%%%%%%%%%%%%%%%%%%%%%%%%%%%%%%%%%%%%%%%%%%%%%%%%%%%%%%%%%%%%%%%%%%%%%%%%%%
Both authors would like to thank Caroline Lasser for fruitful discussions and comments.

%%%%%%%%%%%%%%%%%%%%%%%%%%%%%%%%%%%%%%%%%%%%%%%%%%%%%%%%%%
\section{Symbol Classes, Canonical Transformations \label{s_symbol}}
%%%%%%%%%%%%%%%%%%%%%%%%%%%%%%%%%%%%%%%%%%%%%%%%%%%%%%%%%%

Following the presentation of~\cite{[DimassiSjostrand]} and~\cite{[Martinez]}, we recall the definition of symbol classes.

\begin{defn}[Symbol class]
Let ${\bf d}=(d_j)_{1\leq j\leq J}\in\N^J$, $u(z_1,\ldots,z_J)$ a function of $\Cc^\infty(\R^{d_1}\times\cdots\times\R^{d_J};\C^N)$ and ${\bf m}=(m_j)_{1\leq j\leq J}\in\R^J$. We say that $u$ is a {\bf symbol of class $S[{\bf m};{\bf d}]$} if the following quantities are finite for any $k\geq0$
\begin{equation*}
M^m_k[u]:=\max_{\sum_{j=1}^J\alpha_j=k}\sup_{z_j\in\R^{d_j}}\left|\left(\prod_{j=1}^J\langle z_j\rangle^{-m_j}\partial^{\alpha_j}_{z_j}\right)u(z_1,\ldots,z_J)\right|
\end{equation*}
where $\langle z\rangle:=\sqrt{1+|z|^2}$.

We extend this definition to any $m_j\in\overline{\R}:=\{-\infty\}\cup\R\cup\{+\infty\}$ by setting, for instance with non-finite $m_1$,
\begin{equation*}
S[(+\infty,m_2,\ldots,m_J);{\bf d}]=\bigcup_{m_1\in\R}S[(m_1,\ldots,m_J);{\bf d}]
\end{equation*}
and
\begin{equation*}
S[(-\infty,m_2,\ldots,m_J);{\bf d}]=\bigcap_{m_1\in\R}S[(m_1,\ldots,m_J);{\bf d}]
\end{equation*}
and so on.
\end{defn}

\begin{rem}
$ $
\begin{enumerate}[{\rm(i)}]
\item
$S[{\bf m};{\bf d}]$ is naturally endowed with a Fr\'echet space structure and is increasing with ${\bf m}$ with continuous injection.
\item
Only few values of $m_j$ allow collusion of different $z_j$: if $m=-\infty,0,+\infty$
\begin{equation*}
S[(m,m);(d,d')]=S[m;d+d'].
\end{equation*}
\item
The class $S[(-\infty,\ldots,-\infty);{\bf d}]=S[-\infty;|{\bf d}|]$ coincides with $\Ss(\R^{|{\bf d}|};\C)$ the Schwartz functions on $\R^{|{\bf d}|}$.
\item
We have $S[{\bf m};{\bf d}]S[{\bf m'};{\bf d}]\subset S[{\bf m+m'};{\bf d}]$ with continuous injection.
\item
If $u(z,z')\in S[(m,m');(d,d')]$, then $\langle z\rangle^{-m}u(z,z')\in S[(0,m');(d,d')]$.
\item
To compare with the symbol classes $S^m_{\rho,\delta}$ introduced by H\"ormander, we have $S^m_{0,0}(\R^d\times\R^D)=S[(0,m);(d,D)]$.
\end{enumerate}
\end{rem}

To fix notations, we first recall the definition of a canonical transformation and the link with symplectic matrices.

\begin{defn}[Canonical transformation] \label{def:CanonicalTransform}
Let us consider a smooth diffeomorphism $\kappa(q,p)=(X^\kappa(q,p),\Xi^\kappa(q,p))$ from $T^*\R^d=\R^d\times\R^d$ into itself. We represent its differential by the following Jacobian matrix
\begin{equation} \label{eq:Jacobian}
F^\kappa(q,p)=\left(\begin{array}{cc} X^\kappa_q(q,p)^\dagger & X^\kappa_p(q,p)^\dagger \\ \Xi^\kappa_q(q,p)^\dagger & \Xi^\kappa_p(q,p)^\dagger \end{array}\right).
\end{equation}
$\kappa$ is said to be a {\bf canonical transformation} if $F^\kappa(q,p)$ is symplectic for any $(q,p)$ in $\R^d\times\R^d$ {\it i.e.}
\begin{equation*}
[F^\kappa(q,p)]^\dagger JF^\kappa(q,p)=J \qquad {\rm where} \qquad J:=\left(\begin{array}{cc} 0 & I \\ -I & 0 \end{array}\right).
\end{equation*}
\end{defn}

We specialize here the notion of diffeomorphism of class $\Bb$ as presented by Fujiwara in~\cite{[Fujiwara1]}.

\begin{defn}[Class $\Bb$] \label{def:classB}
A canonical transformation $\kappa$ of $\R^d\times\R^d$ is said to be {\bf of class $\Bb$} if $F^\kappa\in S[0;2d]$. For any $k\geq0$, we set $M^\kappa_k:=M^0_k[F^\kappa]$.
\end{defn}

\begin{rem}
The Hamiltonian flow $\kappa^t$ associated to a subquadratic Hamiltonian function $h(q,p)$ ({\it i.e.} such that the Hessian matrix $\Hess_{(q,p)}h$ is in the class $S[0,2d]$) is of class $\Bb$ for any time $t$ (see~{\rm\cite{[Fujiwara1]}}).
\end{rem}

Once again, for later use, we establish biLipschitzian estimates and properties with respect to composition of canonical transformations.

\begin{lem} \label{invkappa}
The subset of canonical transformations of class $\Bb$ is a subgroup (for composition) of diffeomorphisms of $\R^d\times\R^d$. Moreover, if $\kappa$ is a canonical transformation of class $\Bb$, then there exist two strictly positive constants $c_\kappa$ and $C_\kappa$ such that for any $(q_1,p_1)$ and $(q_2,p_2)$ in $\R^d\times\R^d$
\begin{equation} \label{biLipschitz}
c_\kappa\Vert(q_2,p_2)-(q_1,p_1)\Vert\leq\Vert\kappa(q_2,p_2)-\kappa(q_1,p_1)\Vert\leq C_\kappa\Vert(q_2,p_2)-(q_1,p_1)\Vert.
\end{equation}
\end{lem}

\begin{rem}
\eqref{biLipschitz} is a particular case of the notion of tempered diffeomorphism introduced in {\rm\cite{[Bony]}}. 
\end{rem}

\begin{dem}
For $\kappa$ and $\kappa'$ two canonical transformations of class $\Bb$, we have $F^{\kappa'\circ\kappa}=(F^{\kappa'}\circ\kappa)F^\kappa$ and $F^{\kappa^{-1}}=(F^\kappa\circ\kappa^{-1})^{-1}=-J(F^\kappa\circ\kappa^{-1})^\dagger J$. Thus $M^{\kappa^{-1}}_0=M^\kappa_0$ and $M^{\kappa^{-1}}_k$ (respectively $M_{k}^{\kappa'\circ\kappa}$) are bounded by polynomials in $M^\kappa_l$ (respectively $(M_l^{\kappa}, M_{l'}^{\kappa'})$).
Finally, with $C_\kappa=M^\kappa_0$ and $c_\kappa=[M^{\kappa^{-1}}_0]^{-1}$, \eqref{biLipschitz} directly follows from the Mean Value Inequality applied to $\kappa$ and $\kappa^{-1}$.
\end{dem}

In analogy with the situation of FIOs with real phase, we introduce now the notion of an action associated to a canonical transformation as already suggested in \cite{[Tataru]}. It will play the role usually devoted to a generating function $\underline{S}^\kappa$ associated to $\kappa$ (see Section 5.5 (b) of~{\rm\cite{[Martinez]}}) {\it i.e.} such that
\begin{equation*}
\left(x,\nabla_x\underline{S}^\kappa(x,\eta)\right)=\kappa\left(\nabla_\eta\underline{S}^\kappa(x,\eta),\eta\right).
\end{equation*}

\begin{defn}[Action] \label{def:action}
Let $\kappa=(X^\kappa,\Xi^\kappa)$ be a canonical transformation of $\R^d\times\R^d$. A real-valued function $S^\kappa$ is called an {\bf action associated to $\kappa$} if it fulfills
\begin{equation} \label{eq:action}
S^\kappa_q(q,p)=-p+X^\kappa_q(q,p)\Xi^\kappa(q,p) ,\quad S^\kappa_p(q,p)=X^\kappa_p(q,p)\Xi^\kappa(q,p).
\end{equation}
\end{defn}

\begin{rem} \label{actioninv}
$ $
\begin{enumerate}[{\rm(i)}]
\item
The function $S^\kappa$ always exists and is uniquely defined up to an additive constant. Whenever possible and unambiguous, we will choose the constant of $\kappa={\rm Id}$ so that $S^{\rm Id}=0$. If $\kappa$ and $\kappa'$ are two canonical transformations, then $S^{\kappa'\circ\kappa}=S^{\kappa'}\circ\kappa+S^\kappa$ and $S^{\kappa^{-1}}=-S^\kappa\circ\kappa^{-1}$ where both equalities hold up to an additive constant.
\item
If $\kappa$ is of class $\Bb$ then $S^\kappa$ is $S[2;2d]$, more precisely $\nabla_{(q,p)}S^\kappa$ is $S[1;2d]$.
\end{enumerate}
\end{rem}

From now on, all canonical transformations considered are assumed to be of class $\Bb$ and $\h$ will denote a small parameter such that $0<\h\leq1$.

%%%%%%%%%%%%%%%%%%%%%%%%%%%%%%%%%%%%%%%%%%%%%%%%%%%%%%%%%%
\section{Anti-Wick Calculus and FIOs \label{s_Wick}}
%%%%%%%%%%%%%%%%%%%%%%%%%%%%%%%%%%%%%%%%%%%%%%%%%%%%%%%%%%

We combine here the presentations of Lerner \cite{[Lerner]} and Tataru \cite{[Tataru]} with the additional semiclassical parameter~$\h$ in the spirit of Section 3.4 of~\cite{[Martinez]}. First, we introduce Gaussian wave packets centered in phase space and the FBI transform.

\begin{defn}[FBI transform]
Let $(q,p)\in\R^d\times\R^d$ and $\Theta$ be a complex symmetric (i.e. $\Theta^\dagger=\Theta$) $d\times d$ matrix with positive definite real part $\Re\Theta$. We define
\begin{itemize}
\item
the {\bf normalized coherent state} $g^{\h,\Theta}_{q,p}$ on $\R^d$
\begin{equation*}
g^{\h,\Theta}_{q,p}(y):=\frac{(\det\Re\Theta)^{1/4}}{(\pi\h)^{d/4}}e^{\frac{i}{\h}p\cdot(y-q)}e^{-\frac{\Theta}{2\h}(y-q)\cdot(y-q)},
\end{equation*}
\item
the {\bf FBI transform} of a function $\varphi\in\Ss(\R^d;\C)$
\begin{eqnarray*}
\left[W^\h(\Theta)\varphi\right](q,p) & := & (2\pi\h)^{-d/2}\left\langle g^{\h,\Theta}_{q,p}|\varphi\right\rangle_{L^2_y} \nonumber \\
 & = & \frac{(\det\Re\Theta)^{1/4}}{2^{d/2}(\pi\h)^{3d/4}}\int_{\R^d}e^{-\frac{i}{\h}p\cdot(y-q)}e^{-\frac{\overline{\Theta}}{2\h}(y-q)\cdot(y-q)}\varphi(y)\d y,
\end{eqnarray*}
\item
the {\bf inverse FBI transform} of a function $\Phi\in\Ss(\R^{2d};\C)$
\begin{eqnarray*}
\hspace{-0.5cm}\left[W^\h_{inv}(\Theta)\Phi\right](y) & := & (2\pi\h)^{-d/2}\left\langle\overline{g^{\h,\Theta}_\cdot(y)}\middle|\Phi\right\rangle_{L^2_{(q,p)}} \nonumber \\
 & = & \frac{(\det\Re\Theta)^{1/4}}{2^{d/2}(\pi\h)^{3d/4}}\int_{\R^{2d}}e^{\frac{i}{\h}p\cdot(y-q)}e^{-\frac{\Theta}{2\h}(y-q)\cdot(y-q)}\Phi(q,p)\d q\d p.
\end{eqnarray*}
\end{itemize}
\end{defn}

\begin{rem} \label{rmk:scaling}
Defining $T^\h_d\varphi(y):=\h^{d/4}\varphi(\sqrt{\h}y)$ (which is unitary on $L^2(\R^d;\C)$), we have the scaling formulas
\begin{equation*}
g^{\h,\Theta}_{q,p}=(T^\h_d)^*g^{1,\Theta}_{q/\sqrt{\h},p/\sqrt{\h}},
\end{equation*}
\begin{equation*}
W^\h(\Theta)=(T^\h_{2d})^*W^1(\Theta)T^\h_d, \qquad W^\h_{inv}(\Theta)=(T^\h_d)^*W^1_{inv}(\Theta)T^\h_{2d}.
\end{equation*}
\end{rem}

We recall, with proof, elementary properties of the FBI transform as presented in \cite{[Lerner]}.

\begin{prop}\label{prop:BargmannScontinuous}
If $\varphi\in\Ss(\R^d;\C)$, $\Phi\in\Ss(\R^{2d};\C)$, then $W^\h(\Theta)\varphi\in\Ss(\R^{2d};\C)$ and $W^\h_{inv}(\Theta)\Phi\in\Ss(\R^d;\C)$. Moreover $W^\h(\Theta):\Ss(\R^d;\C)\to\Ss(\R^{2d};\C)$ is continuous, extends by duality to a continuous operator $\Ss'(\R^d;\C)\to\Ss'(\R^{2d};\C)$
~and to a norm preserving map $L^2(\R^d;\C)\to L^2(\R^{2d};\C)$. Finally we have the reconstruction formula
\begin{equation} \label{eq:reconstruction}
\varphi(y)=\frac{1}{(2\pi\h)^{d/2}}\int_{\R^{2d}}\left[W^\h(\Theta)\varphi\right](q,p)g^{\h,\Theta}_{q,p}(y)\d q\d p
\end{equation}
which holds pointwise whenever $\varphi\in\Ss(\R^d;\C)$.
\end{prop}

\begin{rem}
\eqref{eq:reconstruction} corresponds to the fact that $(g^{\h,\Theta}_{q,p})_{(q,p)\in\R^{2d}}$ is an ``overcomplete set of vectors'' of $L^2(\R^d;\C)$ which reads in the ``bra-ket'' notation
\begin{equation*}
\frac{1}{(2\pi\h)^d}\int_{\R^{2d}}\left|g^{\h,\Theta}_{q,p}\right\rangle\left\langle g^{\h,\Theta}_{q,p}\right|\d q\d p={\rm Id}_{L^2(\R^d;\C)}=W^\h_{inv}(\Theta)W^\h(\Theta).
\end{equation*}
However, the composition $W^\h(\Theta)W^\h_{inv}(\Theta)$ is only the orthogonal projection onto the image of $L^2(\R^d;\C)$ under $W^\h(\Theta)$ (see {\rm\cite{[Martinez]}}).
\end{rem}

\begin{dem}
Remark~\ref{rmk:scaling} shows that it is enough to treat the case $\h=1$.

The Schwartz property follows from the fact that $W^1(\Theta)\varphi$ is the partial Fourier transform $(q,y)\to(q,p)$ of a Schwartz function (Schwartz in $y$ and Gaussian in $q$) and an analogous treatment for $W^1_{inv}(\Theta)\Phi$. The isometry property is proven by introducing an extra Gaussian factor to make the integrals absolutely convergent and allow several applications of Fubini's Theorem. $\Vert W^1(\Theta)\varphi\Vert^2$ is the limit $\delta\to0$ of the integral
\begin{eqnarray*}
\lefteqn{\int_{\R^{2d}}e^{-\frac{\delta^2}{2}|p|^2}\overline{[W^1(\Theta)\varphi](q,p)}[W^1(\Theta)\varphi](q,p)\d q\d p} \\
 & = & \int_{\R^{4d}}\frac{e^{-\left[\frac{\delta^2}{2}|p|^2-ip\cdot(y_1-y_2)+\frac{\Theta}{2}(y_1-q)^2+\frac{\overline{\Theta}}{2}(y_2-q)^2\right]}}{2^d\pi^{3d/2}(\det\Re\Theta)^{-1/2}}\overline{\varphi(y_1)}\varphi(y_2)\d y_1\d y_2\d q\d p \\
 & = &  \delta^{-d}\int_{\R^{3d}}\frac{e^{-\left[\frac{\delta^{-2}}{2}|y_1-y_2|^2+\frac{\Theta}{2}(y_1-q)^2+\frac{\overline{\Theta}}{2}(y_2-q)^2\right]}}{2^{d/2}\pi^d(\det\Re\Theta)^{-1/2}}\overline{\varphi(y_1)}\varphi(y_2)\d y_1\d y_2\d q \\
 & = & \int_{\R^{3d}}\frac{e^{-\frac{|w|^2}{2}}e^{-\left[\Re\Theta\underline{q}^2+\frac{\delta^2}{4}\Re\Theta w^2+i\delta\Im\Theta\underline{q}\cdot w\right]}}{2^{d/2}\pi^d(\det\Re\Theta)^{-1/2}}\overline{\varphi\left(y+\frac{\delta}{2}w\right)}\varphi\left(y-\frac{\delta}{2}w\right)\d y\d w\d\underline{q} \\
 & = & \int_{\R^{2d}}\frac{e^{-\frac{|w|^2}{2}}}{(2\pi)^{d/2}}e^{-\frac{\delta^2}{4}\left[\Im\Theta(\Re\Theta)^{-1}\Im\Theta+\Re\Theta\right]w^2}\overline{\varphi\left(y+\frac{\delta}{2}w\right)}\varphi\left(y-\frac{\delta}{2}w\right)\d y\d w
\end{eqnarray*}
with the abuse of notation $\Theta z^2$ for $z\cdot\Theta z$.
Hence
\begin{equation*}
\Vert W^1(\Theta)\varphi\Vert^2=\int_{\R^d}(2\pi)^{-d/2}e^{-\frac{|w|^2}{2}}\d w\int_{\R^d}\overline{\varphi(y)}\varphi(y)\d y=\Vert\varphi\Vert^2.
\end{equation*}
Finally, the reconstruction formula follows from the polarization of the preceding identity:
\begin{eqnarray*}
\langle\psi|\varphi\rangle & = & \langle W^1(\Theta)\psi|W^1(\Theta)\varphi\rangle \\
 & = & (2\pi)^{-d/2}\left\langle\int_{\R^d}\overline{g^{1,\Theta}_{\cdot}(y)}\psi(y)\d y\middle|W^1(\Theta)\varphi\right\rangle \\
 & = & (2\pi)^{-d/2}\int_{\R^{2d}}\int_{\R^d}g^{1,\Theta}_{q,p}(y)\overline{\psi(y)}[W^1(\Theta)\varphi](q,p)\d y\d q\d p \\
 & = & \int_{\R^d}\overline{\psi(y)}\left[(2\pi)^{-d/2}\int_{\R^{2d}}g^{1,\Theta}_{q,p}(y)[W^1(\Theta)\varphi](q,p)\d q\d p\right]\d y.
\end{eqnarray*}
\end{dem}

We define a notion of a FIO with complex phase that generalizes the so-called Anti-Wick quantization of pseudodifferential operators.

\begin{propdef}[Anti-Wick FIO]
Let $u\in L^\infty(\R^d\times\R^d;\C)$, $\kappa$ a canonical transformation and $\Theta^x$, $\Theta^y$ two symmetric $d\times d$ complex matrices with positive definite real part. We define the {\bf semiclassical FIO associated to $\kappa$ with symbol $u$} as the linear operator $\Ii^\h_{AWick}(\kappa;u;\Theta^x,\Theta^y):L^2(\R^d;\C)\to L^2(\R^d;\C)$ such that 
\begin{equation} \label{eq:defFIOWick}
\langle\psi|\Ii^\h_{AWick}(\kappa;u;\Theta^x,\Theta^y)\varphi\rangle_{L^2_y}:=\left\langle[W^\h(\Theta^x)\psi]\circ\kappa\middle|e^{\frac{i}{\h}S^\kappa}u[W^\h(\overline{\Theta^y})\varphi]\right\rangle_{L^2_{(q,p)}}.
\end{equation}
$\Ii^\h_{AWick}(\kappa;u;\Theta^x,\Theta^y)$ is bounded and
\begin{equation} \label{Wickbound}
\Vert \Ii^\h_{AWick}(\kappa;u;\Theta^x,\Theta^y)\Vert_{L^2\to L^2}\leq\Vert u\Vert_{L^\infty}.
\end{equation}
\end{propdef}

\begin{dem}
For any fixed $\varphi\in L^2(\R^d;\C)$, the right-hand side of \eqref{eq:defFIOWick} is a continuous antilinear form in $\psi\in L^2(\R^d;\C)$ with norm bounded by $\Vert u\Vert_{L^\infty}\Vert\varphi\Vert_{L^2}$, so the Riesz representation theorem applies.
\end{dem}

\begin{rem}
$ $
\begin{enumerate}[{\rm(i)}]
\item
We have
\begin{equation*}
\Ii^\h_{AWick}({\rm Id};1;\Theta^x,\Theta^y)=(\det\Re\Theta^x)^{1/4}(\det\Re\Theta^y)^{1/4}\det\left(\frac{\Theta^x+\Theta^y}{2}\right)^{-1/2}{\rm Id}
\end{equation*}
where the choice of the square root is explained in the appendix.
\item
$\Ii^\h_{AWick}({\rm Id};u;I,I)$ is exactly the Anti-Wick quantization of pseudodifferential operators.
\item
The situation with $u=1$ and $\kappa$ linear has been investigated in detail in Section~3.4 of~{\rm\cite{[Martinez]}}.
\item
If $\kappa$ is a canonical transformation of class $\Bb$, $u\in S[0;2d]$ and $\varphi$ is in $\Ss(\R^d;\C)$, then $\Ii^\h_{AWick}(\kappa;u;\Theta^x,\Theta^y)\varphi\in\Ss(\R^d;\C)$ and pointwise
\begin{eqnarray*}
\lefteqn{\left[\Ii^\h_{AWick}(\kappa;u;\Theta^x,\Theta^y)\varphi\right](x)} \\
 & = & \frac{1}{(2\pi\h)^{d/2}}\int_{\R^{2d}}e^{\frac{i}{\h}S^\kappa(q,p)}u(q,p)\left[W^\h(\overline{\Theta^y})\varphi\right](q,p)g^{\h,\Theta^x}_{\kappa(q,p)}(x)\d q\d p.
\end{eqnarray*}
Moreover, if $u\in S[(0,m^p);(d,d)]$ with $m^p<-d$, this last expression equals the absolutely convergent integral
\begin{equation*}
\frac{(\det\Re\Theta^x)^{1/4}(\det\Re\Theta^y)^{1/4}}{2^{-d/2}(2\pi\h)^{3d/2}}\int_{\R^{3d}}e^{\frac{i}{\h}\Phi^\kappa(x,y,q,p;\Theta^x,\Theta^y)}u(q,p)\varphi(y)\d y\d q\d p
\end{equation*}
where
\begin{eqnarray}
\lefteqn{\hspace{-0.7cm} \Phi^\kappa(x,y,q,p;\Theta^x,\Theta^y)=S^\kappa(q,p)+\Xi^\kappa(q,p)\cdot(x-X^\kappa(q,p))-p\cdot(y-q)} \nonumber \\
 & & +\frac{i}{2}(x-X^\kappa(q,p))\cdot\Theta^x(x-X^\kappa(q,p))+\frac{i}{2}(y-q)\cdot\Theta^y(y-q). \label{eq:qphase}
\end{eqnarray}
\item
The result also holds if $S^\kappa$ is not an action associated to $\kappa$. However, the presence of $S^\kappa$ in the oscillating phase is motivated by stationary phase arguments. Indeed, if $u$ is compactly supported the integral kernel of $\Ii^\h_{AWick}(\kappa;u;\Theta^x,\Theta^y)$ is given by
\begin{equation*}
\frac{(\det\Re\Theta^x)^{1/4}(\det\Re\Theta^y)^{1/4}}{2^{-d/2}(2\pi\h)^{3d/2}}\int_{\R^{2d}}e^{\frac{i}{\h}\Phi^\kappa(x,y,q,p;\Theta^x,\Theta^y)}u(q,p)\d q\d p
\end{equation*}
and provides a contribution bigger than $O(\h^\infty)$ if and only if there exists $(q_0,p_0)$ with $u(q_0,p_0)\neq0$, 
\begin{equation*}
\Im\Phi^\kappa(x,y,q_0,p_0;\Theta^x,\Theta^y)=0 \quad {\rm and} \quad \nabla_{(q,p)}\Re\Phi^\kappa(x,y,q_0,p_0;\Theta^x,\Theta^y)=0.
\end{equation*}
The equation on the imaginary part is equivalent to
\begin{equation*}
x-X^\kappa(q_0,p_0)=0 \qquad {\rm and} \qquad y-q_0=0
\end{equation*}
whereas the one on the gradient of the real part reads
\begin{eqnarray*}
S^\kappa_q(q_0,p_0) & = & -p_0+X^\kappa_q(q_0,p_0)\Xi^\kappa(q_0,p_0) \\ 
& & -[\Xi^\kappa_q+X^\kappa_q\Im\Theta^x](q_0,p_0)(x-X^\kappa(q_0,p_0))-\Im\Theta^y(y-q_0) \\
S^\kappa_p(q_0,p_0) & = & X^\kappa_p(q_0,p_0)\Xi^\kappa(q_0,p_0) \\ 
& & -[\Xi^\kappa_p+X^\kappa_p\Im\Theta^x](q_0,p_0)(x-X^\kappa(q_0,p_0))+(y-q_0)
\end{eqnarray*}
whose relation with \eqref{eq:action} is obvious.
\item
With the formal ``bra-ket'' notation, we have
\begin{equation*}
\Ii^\h_{AWick}(\kappa;u;\Theta^x,\Theta^y)=\frac{1}{(2\pi\h)^d}\int_{\R^{2d}}u(q,p)\left|e^{\frac{i}{\h}S^\kappa(q,p)}g^{\h,\Theta^x}_{\kappa(q,p)}\right\rangle\left\langle g^{\h,\overline{\Theta^y}}_{q,p}\right|\d q\d p.
\end{equation*}
\end{enumerate}
\end{rem}

%%%%%%%%%%%%%%%%%%%%%%%%%%%%%%%%%%%%%%%%%%%%%%%%%%%%%%%%%%
\section{FIOs with Complex Phase \label{s_FIO}}
%%%%%%%%%%%%%%%%%%%%%%%%%%%%%%%%%%%%%%%%%%%%%%%%%%%%%%%%%%

Trivial compositions of the FIO of the preceding section with the position operator (either on the left or on the right) show that it could be useful to consider FIOs with symbols depending not only on $(q,p)$ but also on $(x,y)$.

\subsection{Definitions and $\Ss$ Continuity}
%%%%%%%%%%%%%%%%%%%%%%%%%%%%%%%%%%%%%%%%%%%%%%%%%%%%%%%%%%

We define the main object of this article: semiclassical FIOs with quadratic complex phase.

\begin{propdef}[FIO]
Let $\Theta^x$ and $\Theta^y$ be two complex symmetric matrices with positive definite real part. For $u\in S[(+\infty,m^p);(3d,d)]$, $\varphi\in\Ss(\R^d;\C)$ and a positive integer $k>m^p+d$, we define the action of the {\bf semiclassical FIO associated to $\kappa$ with symbol $u$} as the absolutely convergent integral
\begin{eqnarray*}
\lefteqn{[\Ii^\h(\kappa;u;\Theta^x,\Theta^y)\varphi](x):=} \\
 & & \frac{1}{(2\pi\h)^{3d/2}}\int_{\R^{3d}}\hspace{-0.3cm}e^{\frac{i}{\h}\Phi^\kappa(x,y,q,p;\Theta^x,\Theta^y)}(L_y^\dagger)^k\left[u(x,y,q,p)\varphi(y)\right]\d q\d p\d y
\end{eqnarray*}
where $\Phi^\kappa$ is a complex-valued phase function given by \eqref{eq:qphase}, $L_y$ is the first order differential operator
\begin{equation*}
L_y=\frac{1}{1+|\nabla_y\Phi^\kappa(x,y,q,p;\Theta^x,\Theta^y)|^2}\left[1-i\h\nabla_y\overline{\Phi^\kappa(x,y,q,p;\Theta^x,\Theta^y)}\cdot\nabla_y\right]
\end{equation*}
and $L_y^\dagger$ stands for its symmetric given by
\begin{equation*}
\int_{\R^d}v(y)[L_y^\dagger u](y)\d y=\int_{\R^d}[L_yv](y)u(y)\d y.
\end{equation*}
If $m^p<-d$, its integral kernel is given by the absolutely convergent integral
\begin{equation*}
K^\h(\kappa;u;\Theta^x,\Theta^y)(x,y):=\frac{1}{(2\pi\h)^{3d/2}}\int_{\R^{2d}}e^{\frac{i}{\h}\Phi^\kappa(x,y,q,p;\Theta^x,\Theta^y)}u(x,y,q,p)\d q\d p.
\end{equation*}
\end{propdef}

\begin{rem} \label{rmq:Tataru}
$ $
\begin{enumerate}[{\rm(i)}]
\item
As already noticed in {\rm\cite{[AsadaFujiwara]}} or {\rm\cite{[Robert]}}, the following property can be alternatively used as a definition. If $\sigma\in\Ss(\R^d\times\R^d;\C)$ is such that $\sigma(0,0)=1$, we have
\begin{equation} \label{defOscInt}
[\Ii^\h(\kappa;u;\Theta^x,\Theta^y)\varphi](x)=\lim_{\lambda\to+\infty}[\Ii^\h(\kappa;u^\lambda_\sigma;\Theta^x,\Theta^y)\varphi](x) 
\end{equation}
where $u^\lambda_\sigma(x,y,q,p):=\sigma(q/\lambda,p/\lambda)u(x,y,q,p)\in S[(+\infty,-\infty);(2d,2d)]$.
\item
For any $u\in S[+\infty;4d]$, the operator $\Ii^\h(\kappa;u;\Theta^x,\Theta^y)$ is clearly continuous from $\Ss(\R^d;\C)$ into its dual $\Ss'(\R^d;\C)$.
\item
For $(x,y)$-independent symbols $u$, we have
\begin{equation*}
\Ii^\h(\kappa;u;\Theta^x,\Theta^y)=2^{-d/2}(\det\Re\Theta^x\det\Re\Theta^y)^{-1/4}\Ii^\h_{AWick}(\kappa;u;\Theta^x,\Theta^y).
\end{equation*}
In particular, $\Ii^\h({\rm Id};1;\Theta^x,\Theta^y)=\left(\det\left[\Theta^x+\Theta^y\right]^{-1/2}\right){\rm Id}$.
\item
To justify the presence of the action $S^\kappa$ in the oscillating phase, we notice that in general $e^{\frac{i}{\h}S^\kappa}$ does not belong to any symbol class:
\begin{equation*}
|\partial^\alpha(e^{\frac{i}{\h}S^\kappa})|\simeq C\h^{-|\alpha|}|\nabla_{(q,p)}S^\kappa|^{|\alpha|}\simeq C'\h^{-|\alpha|}\langle(q,p)\rangle^{|\alpha|}.
\end{equation*}
\item
With the rescalings
\begin{itemize}
\item
$\kappa^{(\h)}(q,p):=\kappa(\sqrt{\h}q,\sqrt{\h}p)/\sqrt{\h}$ (which preserves the symplectic structure),
\item
$u^{(\h)}(x,y,q,p):=u(\sqrt{\h}x,\sqrt{\h}y,\sqrt{\h}q,\sqrt{\h}p)$,
\end{itemize}
we have $S^{\kappa^{(\h)}}(q,p)=S^\kappa(\sqrt{\h}q,\sqrt{\h}p)/\h$,
\begin{equation*}
\Phi^{\kappa^{(\h)}}(x,y,q,p;\Theta^x,\Theta^y)=\Phi^\kappa(\sqrt{\h}x,\sqrt{\h}y,\sqrt{\h}q,\sqrt{\h}p;\Theta^x,\Theta^y)/\h
\end{equation*}
and
\begin{equation} \label{rescaling}
\Ii^\h(\kappa;u;\Theta^x,\Theta^y)=(T^\h_d)^*\Ii^1(\kappa^{(\h)};u^{(\h)};\Theta^x,\Theta^y)T^\h_d.
\end{equation}
Other rescalings exist with respect to $\Re\Theta^x$ and $\Re\Theta^y$ but they turned out to be useless.
\item
In the non-semiclassical case ($\h=1$), if $\kappa$ is of class $\Bb$ and if the symbol $u\in S[0;3d]$ is independent of $y$, these FIOs are a special case of those introduced in {\rm\cite{[Tataru]}} with symbol
\begin{equation*}
\tilde{u}(x,X^\kappa(q,p),\Xi^\kappa(q,p))=e^{-\frac{1}{2}(x-X^\kappa(q,p))\cdot\Theta^x(x-X^\kappa(q,p))}u(x,q,p).
\end{equation*}
in the symbol class
\begin{equation*}
S_T[0;3d]=\left\{\tilde{u}\in\Cc^\infty(\R^{3d};\C)\middle|\forall\alpha,\beta\geq0,\ \left\Vert(x-X)^\alpha\partial^\beta_{(x,X,\Xi)}\tilde{u}\right\Vert_{L^\infty}<\infty\right\}.
\end{equation*}
The advantage of this last presentation is that, contrary to our situation, the class of FIOs corresponding to this class of symbols form an algebra of bounded operators (for composition) which contains the bounded PDOs. However, as soon as one introduces the semiclassical parameter $\h$, semiclassical expansions of symbols and FIOs are very difficult to obtain in that general setting.
\end{enumerate}
\end{rem}

We now state a result analogous to the situation of pseudodifferential operators which allows clear interpretation of composition of FIOs and composition of a FIO and a pseudodifferential operator. 

\begin{thm} \label{C0S}
If $u\in S[+\infty;4d]$, then $\Ii^\h(\kappa;u;\Theta^x,\Theta^y)$ sends $\Ss(\R^d;\C)$ into itself, is continuous and extends by duality to a continuous operator from $\Ss'(\R^d;\C)$ into itself. More precisely, for finite $m^x$, $m^y$, $m^q$ and $m^p$, the map
\begin{eqnarray*}
S[(m^x,m^y,m^q,m^p);(d,d,d,d)] & \mapsto & (\Ss(\R^d;\C)\to\Ss(\R^d;\C)) \\
u & \to & \Ii^\h(\kappa;u;\Theta^x,\Theta^y)
\end{eqnarray*}
is continuous. Finally, if $u\in S[(+\infty,-\infty);(2d,2d)]$, then $\Ii^\h(\kappa;u;\Theta^x,\Theta^y)$ is smoothing {\it i.e.} $\Ii^\h(\kappa;u;\Theta^x,\Theta^y):\Ss'(\R^d;\C)\to\Ss(\R^d;\C)$.
\end{thm}

Before proving this theorem, we will state and prove interesting intermediate results. The identity~\eqref{rescaling} shows that, in all of those results (including Theorem~\ref{C0S}), it is enough to consider $\h=1$.

We introduce the block matrices
\begin{equation*}
\Sigma_3=\left(\begin{array}{cc} I & 0 \\ 0 & -I \end{array}\right), \qquad
{\mathbf\Theta}^{xy}=\left(\begin{array}{cc} \Theta^x & 0 \\ 0 & \Theta^y \end{array}\right)
\end{equation*}
and compute the derivatives of $\Phi^\kappa$ with respect to the variables $x$, $y$, $q$ and $p$
\begin{equation} \label{derivPhi}
\left(\begin{array}{c} \Phi^\kappa_x \\ \Phi^\kappa_y \\ \Phi^\kappa_q \\ \Phi^\kappa_p \end{array}\right)=
\left(\begin{array}{cc} \Sigma_3 & i{\mathbf\Theta}^{xy} \\ 0 & \W(F^{\kappa}(q,p);\Theta^x,\Theta^y) \end{array}\right)
\left(\begin{array}{c} \Xi^\kappa(q,p) \\ p \\ x-X^\kappa(q,p) \\ y-q \end{array}\right)
\end{equation}
where
\begin{equation*}
\W(F;\Theta^x,\Theta^y):= \left(\begin{array}{c|c} F^\dagger\left(\begin{array}{c} -i\Theta^x \\ I \end{array}\right) & \begin{array}{c} -i\Theta^y \\ -I \end{array}\end{array}\right)=\left(\begin{array}{cc} C^\dagger-iA^\dagger\Theta^x & -i\Theta^y \\ D^\dagger-iB^\dagger\Theta^x & -I \end{array}\right)
\end{equation*}
for a matrix $F$ with block decomposition $\left(\begin{array}{cc} A & B \\ C & D \end{array}\right)$.

We begin by establishing invertibility properties for $\W(F;\Theta^x,\Theta^y)$.

\begin{lem} \label{Winv}
If $F$ is a symplectic matrix, then $\W(F;\Theta^x,\Theta^y)$ is invertible and, if $\kappa$ is a canonical transformation of class $\Bb$ then $\W(F^\kappa(\cdot,\cdot);\Theta^x,\Theta^y)^{-1}$ is in the class $S[0;2d]$.
\end{lem}

\begin{dem}
A straightforward computation shows that
\begin{equation}
\W\left(\Re{\mathbf\Theta}^{xy}\right)^{-1}\W^*=\left[\Lambda\left(\overline{\Theta^y}\right)\right]^\dagger\Lambda\left(\overline{\Theta^y}\right)+\left[\Lambda(\Theta^x)F\right]^\dagger\Lambda(\Theta^x)F \label{Wdefpos}
\end{equation}
where we have introduced the symplectic matrix
\begin{equation*}
\Lambda(\Theta)=\left(\begin{array}{cc} (\Re\Theta)^{1/2} & 0 \\ (\Re\Theta)^{-1/2}\Im\Theta & (\Re\Theta)^{-1/2} \end{array}\right).
\end{equation*}
The matrix on the right-hand side is certainly invertible as the sum of two real symmetric positive definite matrices (because of the invertibility of $F$ for the second), hence the invertibility of $\W(F;\Theta^x,\Theta^y)$.

If $\kappa$ is of class $\Bb$, then $\W(F^\kappa;\Theta^x,\Theta^y)$ is clearly in $S[0;2d]$ so it remains to show that the inverse of the right-hand side in \eqref{Wdefpos} is $S[0;2d]$. As $F^\kappa$ is $S[0;2d]$, the formula of the inverse with minors shows that it is enough to prove a bound from below for the determinant which follows from the following concavity inequality
\begin{equation*}
[\det(A+B)]^{1/d}\geq(\det A)^{1/d}+(\det B)^{1/d}
\end{equation*}
for real symmetric positive matrices $A$ and $B$.
\end{dem}

We state now the best result one can get on the kernel $K^\h$.

\begin{prop} \label{C0Ssmooth}
If $u\in S[-\infty;4d]$, then $K^\h(\kappa;u;\Theta^x,\Theta^y)$ is in $\Ss(\R^d\times\R^d;\C)$. Therefore $\Ii^\h(\kappa;u;\Theta^x,\Theta^y)$ sends $\Ss(\R^d;\C)$ into itself and is continuous. Moreover the map
\begin{equation*}
u\in S[-\infty;4d]\mapsto(\Ii^\h(\kappa;u;\Theta^x,\Theta^y):\Ss(\R^d;\C)\to\Ss(\R^d;\C))
\end{equation*}
is continuous.
\end{prop}

\begin{dem}
By very crude estimates, we get
\begin{eqnarray*}
\Vert K^1(\kappa;u;\Theta^x,\Theta^y)\Vert_{L^\infty_{(x,y)}} & \leq & \frac{1}{(2\pi)^{3d/2}}\Vert u\Vert_{L^\infty_{(x,y)}L^1_{(q,p)}} \\
 & \leq & \frac{1}{(2\pi)^{3d/2}}\Vert \langle(q,p)\rangle^{2d+1}u\Vert_{L^\infty_{(x,y,q,p)}}\int_{\R^{2d}}\frac{\d z}{\langle z\rangle^{2d+1}}.
\end{eqnarray*}
Moreover, we have $(x,y)^\alpha\partial^\beta_{(x,y)}\left(e^{i\Phi^\kappa}u\right)=e^{i\Phi^\kappa}v_{\alpha\beta}(\kappa;u;\Theta^x,\Theta^y)$ where
\begin{equation} \label{ProdDerivExp}
v_{\alpha\beta}(\kappa;u;\Theta^x,\Theta^y)=\sum_{\gamma_0+\gamma=\beta}\binom{\beta}{\gamma}\sum_{k=1}^{|\gamma|}i^k\sum_{\substack{\gamma_1+\cdots+\gamma_k=\gamma \\ |\gamma_j|\geq1}}(x,y)^\alpha\partial^{\gamma_0}_{(x,y)}u\prod_{j=1}^k\partial^{\gamma_j}_{(x,y)}\Phi^\kappa.
\end{equation}
From \eqref{derivPhi}, we see that $\Phi^\kappa_{(x,y)}$ is $S[1;4d]$ so that $v_{\alpha\beta}(\kappa;u;\Theta^x,\Theta^y)$ is $S[-\infty;4d]$. Thus, by dominated convergence, $K^1(\kappa;u;\Theta^x,\Theta^y)$ is smooth and 
\begin{equation*}
(x,y)^\alpha\partial^\beta_{(x,y)}K^1(\kappa;u;\Theta^x,\Theta^y)=K^1(\kappa;v_{\alpha\beta}(\kappa;u;\Theta^x,\Theta^y);\Theta^x,\Theta^y)
\end{equation*}
with
\begin{eqnarray}
\lefteqn{\left\Vert(x,y)^\alpha\partial^\beta_{(x,y)}K^1(\kappa;u;\Theta^x,\Theta^y)\right\Vert_{L^\infty_{(x,y)}}\leq C} \nonumber \\
& & \hspace{1cm}\times\sum_{|\gamma|\leq|\beta|}\left\Vert \langle(q,p)\rangle^{2d+1}\langle(x,y)\rangle^{|\alpha|}\partial^\gamma_{(x,y)}u\langle(x,y,q,p)\rangle^{|\beta|-|\gamma|}\right\Vert_{L^\infty_{(x,y,q,p)}}. \label{poly_xy}
\end{eqnarray}
\end{dem}

Estimate \eqref{poly_xy} requires fast decay in $(x,y)$ for the symbol $u$ but in fact, one can drop any polynomial growth in $(x,y)$ ``by hand''.

\begin{prop} \label{C0Shalfsmooth}
If $u\in S[(+\infty,-\infty);(2d,2d)]$, then the corresponding kernel $K^\h(\kappa;u;\Theta^x,\Theta^y)$ is in $\Ss(\R^d\times\R^d;\C)$. Therefore $\Ii^\h(\kappa;u;\Theta^x,\Theta^y)$ sends $\Ss(\R^d;\C)$ into itself and is continuous. Moreover, for finite $m^x$ and $m^y$, the map
\begin{equation*}
u\in S[(m^x,m^y,-\infty);(d,d,2d)]\mapsto(\Ii^\h(\kappa;u;\Theta^x,\Theta^y):\Ss(\R^d;\C)\to\Ss(\R^d;\C))
\end{equation*}
is continuous.
\end{prop}

\begin{dem}
Let us assume first that $u\in S[(0,0,-\infty);(d,d,2d)]$. Following the lines of the preceding proposition, we have
\begin{equation*}
v_{\alpha\beta}(\kappa;u;\Theta^x,\Theta^y)=\sum_{\gamma\leq\beta}P_{\alpha\beta}^\gamma[\Theta^x,\Theta^y](x,y,p,\Xi^\kappa(q,p),x-X^\kappa(q,p),y-q)\partial^\gamma_{(x,y)}u
\end{equation*}
where $P_{\alpha\beta}^\gamma[\Theta^x,\Theta^y]$ is polynomial in its variables of degree $|\alpha|+|\beta|-|\gamma|$ at most. By an exact Taylor expansion, we get
\begin{eqnarray*}
\lefteqn{P_{\alpha\beta}^\gamma[\Theta^x,\Theta^y](x,y,p,\Xi^\kappa(q,p),x-X^\kappa(q,p),y-q)} \\
 & & \hspace{2cm}=\sum_{|\delta|\leq|\alpha|+|\beta|-|\gamma|}Q_{\alpha\beta}^{\gamma\delta}[\kappa,\Theta^x,\Theta^y](q,p)(x-X^\kappa(q,p),y-q)^\delta
\end{eqnarray*}
where $Q_{\alpha\beta}^{\gamma\delta}[\kappa,\Theta^x,\Theta^y]$ is $S[|\alpha|+|\beta|-|\gamma|-|\delta|,2d]$. Now, we state and prove the lemma that will allow us to transfer the polynomial growth in $(x,y)$ into $(q,p)$.

\begin{lem} \label{push}
Let $u\in S[(m^x,m^y,m^q,m^p);(d,d,d,d)]$ with $m^p<-d$ and $V$ a constant vector of $\C^d\times\C^d$, then
\begin{equation} \label{eq:IPP}
K^\h\left(\kappa;V\cdot\left(\begin{array}{c} x-X^\kappa(q,p) \\ y-q \end{array}\right)u;\Theta^x,\Theta^y\right)=i\h K^\h\left(\kappa;L(\kappa;\Theta^x,\Theta^y;V)u;\Theta^x,\Theta^y\right)
\end{equation}
where
\begin{equation} \label{IPP}
[L(\kappa;\Theta^x,\Theta^y;V)u](x,y,q,p):=\div_{(q,p)}\left[u(x,y,q,p){\W(F^\kappa(q,p);\Theta^x,\Theta^y)^\dagger}^{-1}V\right]
\end{equation}
and $L(\kappa;\Theta^x,\Theta^y;V)$ is continuous from $S[(m^x,m^y,m^q,m^p);(d,d,d,d)]$ into itself.
\end{lem}

\begin{dem}
By \eqref{derivPhi}, we have
\begin{equation} \label{forIPP}
\left(\begin{array}{c} x-X^\kappa(q,p) \\ y-q \end{array}\right)e^{\frac{i}{\h}\Phi^\kappa}=-i\h \W(F^\kappa(q,p);\Theta^x,\Theta^y)^{-1}\left(\begin{array}{c} \nabla_q \\ \nabla_p \end{array}\right)e^{\frac{i}{\h}\Phi^\kappa},
\end{equation}
hence the equality integrating by parts and the continuity by Lemma~\ref{Winv}.
\end{dem}

By iterative applications of this Lemma, we have
\begin{eqnarray*}
\lefteqn{\hspace{-0.5cm}K^1\left(\kappa;(x-X^\kappa(q,p),y-q)^\delta Q_{\alpha\beta}^{\gamma\delta}[\kappa,\Theta^x,\Theta^y]\partial^\gamma_{(x,y)}u;\Theta^x,\Theta^y\right)} \\
 & & \hspace{3cm}=K^1\left(\kappa;\sum_{|\mu|\leq\delta}R_{\alpha\beta}^{\gamma\delta\mu}[\kappa,\Theta^x,\Theta^y]\partial^\mu_{(q,p)}\partial^\gamma_{(x,y)}u;\Theta^x,\Theta^y\right)
\end{eqnarray*}
where $R_{\alpha\beta}^{\gamma\delta\mu}[\kappa,\Theta^x,\Theta^y]$ is $S[|\alpha|+|\beta|-|\gamma|-|\delta|,2d]$. Hence
\begin{eqnarray*}
\lefteqn{\Vert(x,y)^\alpha\partial^\beta_{(x,y)}K^1(\kappa;u;\Theta^x,\Theta^y)\Vert_{L^\infty_{(x,y)}}\leq C} \\
& & \hspace{1cm}\times\sum_{\substack{|\gamma|\leq|\beta| \\ |\delta|\leq|\alpha|+|\beta|-|\gamma| \\ |\mu|\leq\delta}}\Vert \langle(q,p)\rangle^{2d+1+|\alpha|+|\beta|-|\gamma|-|\delta|}\partial^\mu_{(q,p)}\partial^\gamma_{(x,y)}u\Vert_{L^\infty_{(x,y,q,p)}}.
\end{eqnarray*}

The continuous injection induces the result for any negative $m^x$ and $m^y$. To get it for any $m^x$ and $m^y$, we will prove it for $m^x=m^y=2m$ with $m$ a positive integer. If $u\in S[(2m,2m,-\infty);(d,d,2d)]$ then $v:=\langle x\rangle^{-2m}\langle y\rangle^{-2m}u$ is $S[(0,0,-\infty);(d,d,2d)]$ and, by iterative applications of Lemma \ref{push} as before,
\begin{eqnarray*}
K^1(\kappa;u;\Theta^x,\Theta^y) & = & K^1(\kappa;\langle x\rangle^{2m}\langle y\rangle^{2m}v;\Theta^x,\Theta^y) \\
 & = & \sum_{\substack{|\delta|\leq4m \\ |\mu|\leq|\delta|}}K^1(\kappa;R_m^{\delta\mu}[\kappa,\Theta^x,\Theta^y]\partial^\mu_{(q,p)}v;\Theta^x,\Theta^y)
\end{eqnarray*}
with $R_m^{\delta\mu}[\kappa,\Theta^x,\Theta^y]\in S[4m-|\delta|;2d]$ and we are back to the preceding case.
\end{dem}

We have now reached the best we can do as far as the kernel is concerned. The next step consists in compensating the possible polynomial growth in $(q,p)$ of the symbol $u$ by the fact that the function $\varphi$ on which we apply the operator has fast decay.

{\bf Proof of Theorem~\ref{C0S}:}
Rereading \eqref{ProdDerivExp}, we see that, for $\varphi\in\Ss(\R^d;\C)$ and $u\in S[(+\infty,-\infty);(2d,2d)]$,
\begin{eqnarray}
\lefteqn{(x^\alpha\partial^\beta_x)[\Ii^1(\kappa;u;\Theta^x,\Theta^y)\varphi]=} \label{ProdDerivOp} \\
 & & \sum_{\gamma_0+\gamma=\beta}\binom{\beta}{\gamma}\sum_{k=1}^{|\gamma|}i^k\sum_{\substack{\gamma_1+\cdots+\gamma_k=\gamma \\ |\gamma_j|\geq1}}\Ii^1\left(\kappa;x^\alpha\partial^{\gamma_0}_xu\prod_{j=1}^k\partial^{\gamma_j}_x\Phi^\kappa;\Theta^x,\Theta^y\right)\varphi. \nonumber
\end{eqnarray}
Thus, if $u\in S[+\infty;4d]$ and $\varphi\in\Ss(\R^d;\C)$, introducing $\sigma$ and passing to the limit $\lambda\to+\infty$, we get that $\Ii^\h(\kappa;u;\Theta^x,\Theta^y)\varphi$ is smooth and \eqref{ProdDerivOp} holds. It remains to estimate the right-hand side of \eqref{ProdDerivOp} in norm $L^\infty$.

We assume that $u\in S[(2m,2m,2m);(d,d,2d)]$ with $m$ a non-negative integer. As $\partial^{\gamma_j}_x\Phi^\kappa$ is polynomial with degree at most~$1$ in $(\Xi^\kappa(q,p),x-X^\kappa(q,p),y-q)$ with coefficients in $S[0;2d]$ and using exact Taylor expansion in $x=X^\kappa(q,p)$ for $x^\alpha\langle x\rangle^{2m}$, we get
\begin{eqnarray*}
\lefteqn{x^\alpha\partial^{\gamma_0}_xu\prod_{j=1}^k\partial^{\gamma_j}_x\Phi^\kappa=} \\
& & \hspace{-0.8cm} \sum_{|\delta|\leq|\alpha|+2m+k} \hspace{-0.5cm} P^{\alpha\delta m}_{\gamma_1\cdots\gamma_k}[\Theta^x,\Theta^y](X^\kappa(q,p),\Xi^\kappa(q,p),y-q)(x-X^\kappa(q,p))^\delta[\langle x\rangle^{-2m}\partial^{\gamma_0}_xu]
\end{eqnarray*}
where $P^{\alpha\delta m}_{\gamma_1\cdots\gamma_k}[\Theta^x,\Theta^y]$ is polynomial in its variables of degree $|\alpha|+2m+k-|\delta|$ at most with coefficients depending only on $\Theta^x$ and $\Theta^y$. We now generalize Lemma \ref{push} to the operator case.

\begin{lem} \label{push2}
Let $u\in S[+\infty;4d]$ and $V$ a constant vector of $\C^d\times\C^d\times\C^d\times\C^d$, then, for any $\varphi\in\Ss(\R^d;\C)$,
\begin{eqnarray*}
\lefteqn{\Ii^\h\left(\kappa;V\cdot\left(\begin{array}{c} \Xi^\kappa(q,p) \\ p \\ x-X^\kappa(q,p) \\ y-q \end{array}\right)u;\Theta^x,\Theta^y\right)\varphi} \\
 & = & i\h \Ii^\h(\kappa;L'(\kappa;\Theta^x,\Theta^y;V)u;\Theta^x,\Theta^y)\varphi-i\h \Ii^\h(\kappa;u;\Theta^x,\Theta^y)[V^y\cdot\nabla_y\varphi]
\end{eqnarray*}
where
\begin{eqnarray} \label{IPP2}
L'(\kappa;\Theta^x,\Theta^y;V)u & := & L\left(\kappa;\Theta^x,\Theta^y;\left(\begin{array}{c} V^q \\ V^p \end{array}\right)-i{\mathbf\Theta}^{xy}\left(\begin{array}{c} V^x \\ -V^y \end{array}\right)\right)u \\
 & & +[V^x\cdot\nabla_xu-V^y\cdot\nabla_yu] \nonumber
\end{eqnarray}
and $L'(\kappa;\Theta^x,\Theta^y;V)$ is continuous from $S[(m^x,m^y,m^q,m^p);(d,d,d,d)]$ into itself.
\end{lem}

\begin{rem}
In particular, the identity~\eqref{eq:IPP} also holds for $u\in S[+\infty;4d]$ but in a distributional sense.
\end{rem}

\begin{dem}
By \eqref{derivPhi}, we have
\begin{eqnarray}
\lefteqn{\left(\begin{array}{c} \Xi^\kappa(q,p) \\ p \\ x-X^\kappa(q,p) \\ y-q \end{array}\right)e^{\frac{i}{\h}\Phi^\kappa}=} \label{forIPP2} \\
 & & -i\h\left(\begin{array}{cc} \Sigma_3 & -i\Sigma_3{\mathbf\Theta}^{xy} \\ 0 & I \end{array}\right)\left(\begin{array}{cc} I & 0 \\ 0 & \W(F^\kappa(q,p);\Theta^x,\Theta^y)^{-1} \end{array}\right)\left(\begin{array}{c} \nabla_x \\ \nabla_y \\ \nabla_q \\ \nabla_p \end{array}\right)e^{\frac{i}{\h}\Phi^\kappa}, \nonumber
\end{eqnarray}
hence the equality for $u\in S[(m^x,m^y,-\infty);(d,d,2d)]$ integrating by parts.

We deduce it then for $u\in S[(m^x,m^y,m^q,m^p);(d,d,d,d)]$ passing to the limit $\lambda\to+\infty$ on both sides and using that
\begin{eqnarray*}
\lefteqn{L'(\kappa;\Theta^x,\Theta^y;V)u^\lambda_\sigma=[L'(\kappa;\Theta^x,\Theta^y;V)u]^\lambda_\sigma} \\
 & & \hspace{1.5cm}+\lambda^{-1}u{\W^\dagger}^{-1}\left[\left(\begin{array}{c} V^q \\ V^p \end{array}\right)-i{\mathbf\Theta}^{xy}\left(\begin{array}{c} V^x \\ -V^y \end{array}\right)\right]\cdot(\nabla_{(q,p)}\sigma)\left(\frac{q}{\lambda},\frac{p}{\lambda}\right).
\end{eqnarray*}
The continuity follows because of Lemma~\ref{Winv}.
\end{dem}

Lemma \ref{push2} has the particular case, for $V$ a constant vector of $\C^d$,
\begin{equation*}
\Ii^1(\kappa;V\cdot(x-X^\kappa(q,p))u;\Theta^x,\Theta^y)=i\Ii^1\left(\kappa;\div_{(q,p)}\left[u\W^{-1}\left(\begin{array}{c} V \\ 0 \end{array}\right)\right];\Theta^x,\Theta^y\right)
\end{equation*}
which, by iterative applications, induces that
\begin{eqnarray*}
\lefteqn{\Ii^1\left(\kappa;x^\alpha\partial^{\gamma_0}_xu\prod_{j=1}^k\partial^{\gamma_j}_x\Phi^\kappa;\Theta^x,\Theta^y\right)=\sum_{\substack{|\delta|\leq|\alpha|+2m+k \\ |\mu|\leq|\delta|}}} \\
& & \hspace{-0.5cm}\Ii^1\left(\kappa;Q^{\alpha\delta\mu m}_{\gamma_1\cdots\gamma_k}[\Theta^x,\Theta^y](X^\kappa(q,p),\Xi^\kappa(q,p),y-q)\partial^\mu_{(q,p)}[\langle x\rangle^{-2m}\partial^{\gamma_0}_xu];\Theta^x,\Theta^y\right)
\end{eqnarray*}
where $Q^{\alpha\delta\mu m}_{\gamma_1\cdots\gamma_k}[\Theta^x,\Theta^y]$ is polynomial in its variables of degree $|\alpha|+2m+k-|\delta|$ at most with coefficients depending on $(q,p)$ (in $S[0;2d]$), $\Theta^x$ and $\Theta^y$.

By a trivial Taylor expansion in $(q,p)=(y,0)$, we have
\begin{equation*}
\left(\begin{array}{c} X^\kappa(q,p) \\ \Xi^\kappa(q,p) \end{array}\right)=\left(\begin{array}{c} X^\kappa(y,0) \\ \Xi^\kappa(y,0) \end{array}\right)+G^\kappa(y,q,p)\left(\begin{array}{c} q-y \\ p \end{array}\right)
\end{equation*}
where
\begin{equation*}
G^\kappa(y,q,p):=\int^1_0F^\kappa(y+\tau(q-y),\tau p)\d\tau\in S[0;3d]
\end{equation*}
as $\kappa$ is of class $\Bb$. Thus,
\begin{eqnarray*}
\lefteqn{Q^{\alpha\delta\mu m}_{\gamma_1\cdots\gamma_k}[\Theta^x,\Theta^y](X^\kappa(q,p),\Xi^\kappa(q,p),y-q)} \\
 & & \hspace{3.5cm}=R^{\alpha\delta\mu m}_{\gamma_1\cdots\gamma_k}[\Theta^x,\Theta^y](X^\kappa(y,0),\Xi^\kappa(y,0),y-q,p)
\end{eqnarray*}
where $R^{\alpha\delta\mu m}_{\gamma_1\cdots\gamma_k}[\Theta^x,\Theta^y]$ is polynomial in its variables of degree $|\alpha|+2m+k-|\delta|$ at most with coefficients depending on $(y,q,p)$ (in $S[0;3d]$), $\Theta^x$ and $\Theta^y$.

Lemma \ref{push2} has the second particular case, for $V^y$ and $V^p$ two constant vectors of $\C^d$,
\begin{eqnarray*}
\lefteqn{\Ii^1(\kappa;[V^y\cdot p+V^p\cdot(y-q)]u;\Theta^x,\Theta^y)\varphi=} \\
 & & \hspace{-0.3cm}i\Ii^1\left(\kappa;\div_{(q,p)}\left(u\W^{-1}\left[\left(\begin{array}{c} 0 \\ V^p \end{array}\right)+i{\mathbf\Theta}^{xy}\left(\begin{array}{c} 0 \\ V^y \end{array}\right)\right]\right)-V^y\cdot\nabla_yu;\Theta^x,\Theta^y\right)\varphi \\
 & & -i\Ii^1(\kappa;u;\Theta^x,\Theta^y)[V^y\cdot\nabla_y\varphi]
\end{eqnarray*}
which, by iterative applications, induces that
\begin{eqnarray*}
\lefteqn{\Ii^1\left(\kappa;x^\alpha\partial^{\gamma_0}_xu\prod_{j=1}^k\partial^{\gamma_j}_x\Phi^\kappa;\Theta^x,\Theta^y\right)\varphi=\sum_{\substack{|\delta|\leq|\alpha|+2m+k \\ |\mu|\leq|\delta| \\ |\mu_0|+|\mu_1|\leq|\alpha|+8m+k-|\delta|+2d+2}}} \\
 & & \hspace{-0.7cm}\Ii^1\left(\kappa;R^{\alpha\delta\mu\mu_0\mu_1m}_{\gamma_1\cdots\gamma_k}[\Theta^x,\Theta^y]\partial^{\mu_0}_{(y,q,p)}\left[\frac{\partial^\mu_{(q,p)}\partial^{\gamma_0}_xu}{(\langle x\rangle\langle y\rangle\langle q\rangle\langle p\rangle)^{2m}\langle(q,p)\rangle^{2d+2}}\right];\Theta^x,\Theta^y\right)\partial^{\mu_1}_y\varphi
\end{eqnarray*}
where $R^{\alpha\delta\mu\mu_0\mu_1m}_{\gamma_1\cdots\gamma_k}[\Theta^x,\Theta^y]$ is polynomial in $(y,X^\kappa(y,0),\Xi^\kappa(y,0))$ of degree at most $|\alpha|+8m+k-|\delta|+2d+2$ with coefficients depending on $(y,q,p)$ (in $S[0;3d]$), $\Theta^x$ and $\Theta^y$.

The last step is to observe that $y\to\kappa(y,0)$ is $S[1;d]$ so that
\begin{eqnarray*}
\lefteqn{\Vert(x^\alpha\partial^\beta_x)[\Ii^1(\kappa;u;\Theta^x,\Theta^y)\varphi]\Vert_{L^\infty}\leq C\int_{\R^{2d}}\frac{\d q\d p}{\langle(q,p)\rangle^{2d+2}}\int_{\R^d}\frac{\d y}{\langle y\rangle^{d+1}}} \\
 & & \times\sum\left\Vert\partial^{\mu_0}_{(y,q,p)}\left[\frac{\partial^\mu_{(q,p)}\partial^{\gamma_0}_xu}{(\langle x\rangle\langle y\rangle\langle q\rangle\langle p\rangle)^{2m}}\right]\right\Vert_{L^\infty}\left\Vert\langle y\rangle^{|\alpha|+6m+k-|\delta|+3d+3}\partial_y^{\mu_1}\varphi\right\Vert_{L^\infty}%.
\end{eqnarray*}
where the indices of summation can easily be deduced from above.
\hspace {\stretch{1} }$\Box$

\begin{rem}
The opportunity of stating results with minimal regularity of the type
\begin{equation*}
\Ii^\eps(\kappa;u;\Theta^x,\Theta^y):\Cc^{k_1}_c(\R^d;\C)\to\Cc^{k_2}(\R^d;\C)\cap L^2(\R^d;\C)
\end{equation*}
for symbols $u$ with $M^m_k[u]<\infty$ only for $k\leq k_3$ and canonical transformation $\kappa$ with $M^\kappa_k<\infty$ only for $k\leq k_4$ (with appropriate $(k_1,k_2,k_3,k_4)$) is left to the reader.
\end{rem}

%%%%%%%%%%%%%%%%%%%%%%%%%%%%%%%%%%%%%%%%%%%%%%%%%%%%%%%%%%
\subsection{Formal Adjoint}
%%%%%%%%%%%%%%%%%%%%%%%%%%%%%%%%%%%%%%%%%%%%%%%%%%%%%%%%%%

As they will be a useful tool for the proof of the main theorem, we prove some abstract results about the behaviour of FIOs with respect to adjunction.

\begin{defn}[Formal adjoint]
Let $\Ii,\Ii':\Ss(\R^d;\C)\to\Ss(\R^d;\C)$ be two linear operators. We say that $\Ii'$ is a {\bf formal adjoint} of $\Ii$ if, for any $\varphi,\psi\in\Ss(\R^d;\C)$, we have
\begin{equation*}
\langle\Ii\varphi|\psi\rangle_{L^2(\R^d;\C)}=\langle\varphi|\Ii'\psi\rangle_{L^2(\R^d;\C)}.
\end{equation*}
In that situation, we use the notation $\Ii^{(*)}$ for $\Ii'$.
\end{defn}

\begin{rem}
$ $
\begin{enumerate}[{\rm(i)}]
\item
When it exists, the formal adjoint is necessarily unique because $\Ss(\R^d;\C)$ is dense in $L^2(\R^d;\C)$.
\item
Both the denomination and notation come from the following statement: if $\Ii$ can be extended to a (necessarily unique) linear bounded operator $\overline{\Ii}:L^2(\R^d;\C)\to L^2(\R^d;\C)$, then so does $\Ii^{(*)}$ and $\overline{\Ii}^*=\overline{\Ii^{(*)}}$ where the left-hand side is the usual adjoint of linear bounded operators.
\item
For two linear operators $\Ii,\Ii':\Ss(\R^d;\C)\to\Ss(\R^d;\C)$, $(\Ii^{(*)})^{(*)}=\Ii$ and $(\Ii\Ii')^{(*)}={\Ii'}^{(*)}\Ii^{(*)}$.
\item
If $\Ii^{(*)}\Ii$ can be extended to a bounded operator $L^2(\R^d;\C)\to L^2(\R^d;\C)$, then so do $\Ii$ and $\Ii^{(*)}$, moreover we have $\Vert\Ii\Vert^2=\Vert\Ii^{(*)}\Vert^2=\Vert\Ii^{(*)}\Ii\Vert$.
\end{enumerate}
\end{rem}

\begin{prop}
Let $u\in S[+\infty;4d]$ and $\kappa$ be a canonical transformation of class $\Bb$, then, as operators $\Ss(\R^d;\C)\to\Ss(\R^d;\C)$,
\begin{equation*}
\Ii^\h(\kappa;u;\Theta^x,\Theta^y)^{(*)}=e^{\frac{i}{\h}C}\Ii^\h(\kappa^{-1};u^\kappa;\overline{\Theta^y},\overline{\Theta^x})
\end{equation*}
where $u^\kappa(x,y,q,p)=\overline{u(y,x,X^{\kappa^{-1}}(q,p),\Xi^{\kappa^{-1}}(q,p))}$ and $C$ is a constant that depends on the actions associated to $\kappa$ and $\kappa^{-1}$.
\end{prop}

\begin{dem}
We restrict to $u\in S[(+\infty,-\infty);(2d,2d)]$ to show the strategy at the kernel level in a compact manner. The proof easily extends to the operator level introducing as many $L_y^\dagger$ as necessary to make the integral with respect to $p$ absolutely convergent. The kernel of $\Ii^\h(\kappa;u;\Theta^x,\Theta^y)^{(*)}$ is
\begin{eqnarray*}
\lefteqn{\overline{K^\h(\kappa;u;\Theta^x,\Theta^y)(y,x)}} \\
 & = & \frac{1}{(2\pi\h)^{3d/2}}\int_{\R^{2d}}e^{-\frac{i}{\h}\overline{\Phi^\kappa(y,x,q,p;\Theta^x,\Theta^y)}}\overline{u(y,x,q,p)}\d q\d p \\
 & = & \frac{1}{(2\pi\h)^{3d/2}}\int_{\R^{2d}}e^{-\frac{i}{\h}\overline{\Phi^\kappa(y,x,X^{\kappa^{-1}}(q',p'),\Xi^{\kappa^{-1}}(q',p');\Theta^x,\Theta^y)}}u^\kappa(x,y,q',p')\d q'\d p'
\end{eqnarray*}
with the symplectic change of variables $(q',p')=\kappa(q,p)$. To conclude, it remains to show that
\begin{equation*}
-\overline{\Phi^\kappa(y,x,X^{\kappa^{-1}}(q',p'),\Xi^{\kappa^{-1}}(q',p');\Theta^x,\Theta^y)}=\Phi^{\kappa^{-1}}(x,y,q',p';\overline{\Theta^y},\overline{\Theta^x})
\end{equation*}
up to an additive constant, which follows from a straightforward computation and Remark \ref{actioninv}.
\end{dem}

\begin{cor} \label{II*eqI*I}
Let $u_1,u_2\in S[+\infty;4d]$ and $\kappa_1,\kappa_2$ be two canonical transformations of class $\Bb$, then
\begin{eqnarray*}
\lefteqn{\Ii^\h(\kappa_1;u_1;\Theta^x_1,\Theta^y_1)\Ii^\h(\kappa_2;u_2;\Theta^x_2,\Theta^y_2)^{(*)}=} \\
 & & \hspace{3cm}e^{\frac{i}{\h}C}\Ii^\h(\kappa^{-1}_1;u^{\kappa_1}_1;\overline{\Theta^y_1},\overline{\Theta^x_1})^{(*)}\Ii^\h(\kappa^{-1}_2;u^{\kappa_2}_2;\overline{\Theta^y_2},\overline{\Theta^x_2})
\end{eqnarray*}
where $C$ is a constant depending on the four actions involved.
\end{cor}

%%%%%%%%%%%%%%%%%%%%%%%%%%%%%%%%%%%%%%%%%%%%%%%%%%%%%%%%%%
\subsection{$L^2$ Continuity}
%%%%%%%%%%%%%%%%%%%%%%%%%%%%%%%%%%%%%%%%%%%%%%%%%%%%%%%%%%

In this last section, we will prove an $L^2$-boundedness result for our FIOs analogous to the Calder\'on-Vaillancourt Theorem for pseudodifferential operators. We assume that $0<\lambda^xI\leq\Re\Theta^x\leq\gamma^xI$ as quadratic forms with analogous inequalities for $\Re\Theta^y$ (one can take $\gamma^x=\Vert\Re\Theta^x\Vert$ and $\lambda^x=\Vert(\Re\Theta^x)^{-1}\Vert^{-1}$).

\begin{thm} \label{theo:L2bound}
Let $u\in S[0;4d]$ be a symbol, then $\Ii^\h(\kappa;u;\Theta^x,\Theta^y)$ can be extended in a unique way to a linear bounded operator $L^2(\R^d;\C)\to L^2(\R^d;\C)$ and there exists $C>0$ such that
\begin{eqnarray}
\lefteqn{\left\Vert\Ii^\h(\kappa;u;\Theta^x,\Theta^y)\right\Vert_{L^2\to L^2}\leq C} \nonumber \\
 & & \times\left(1+\frac{1}{[\min(1,\lambda^x,\lambda^y)\eta^2_{[\kappa,\Theta^x,\Theta^y]}]^{(4d+1)/4}}\right)\frac{\Vert u\Vert_{W^{4d+1,\infty}_{(x,y)}L^\infty_{(q,p)}}}{(\det\Re\Theta^x\det\Re\Theta^y)^{1/4}} \label{eq:full}
\end{eqnarray}
where $\eta_{[\kappa,\Theta^x,\Theta^y]}$ is defined by~\eqref{eq:eta} and
\begin{equation*}
\Vert w\Vert_{W^{4d+1,\infty}_{(x,y)}L^\infty_{(q,p)}}:=\sum_{|\alpha|\leq4d+1}\Vert\partial^\alpha_{(x,y)}w\Vert_{L^\infty}.
\end{equation*}
In the special case where $u\in S[0;2d]$ is independent of $(x,y)$, restating \eqref{Wickbound}, we have
\begin{equation} \label{eq:corfull}
\left\Vert \Ii^\h(\kappa;u;\Theta^x,\Theta^y)\right\Vert_{L^2\to L^2}\leq 2^{-d/2}\frac{\Vert u\Vert_{L^\infty}}{(\det\Re\Theta^x\det\Re\Theta^y)^{1/4}}.
\end{equation}
\end{thm}

\begin{rem} \label{rmk:scaleinv}
With respect to the semiclassical parameter $\h$, \eqref{eq:full} has the same type of scale invariance as Calder\'on-Vaillancourt Theorem: namely the result for $\h=1$ induces a result for $\h<1$ a little bit stronger than the one stated as
\begin{equation*}
\Vert\partial^\alpha_{(x,y)}u^{(\h)}\Vert_{L^\infty}=\h^{|\alpha|/2}\Vert\partial^\alpha_{(x,y)}u\Vert_{L^\infty}.
\end{equation*}
\end{rem}

As for the case of $\Ss$ continuity, the proof will proceed by steps of increasing difficulty. Each step relies on the well-known Schur's Lemma for integral operators.

\begin{lem}[Schur]
If $A\varphi(x)=\int_{\R^d}K(x,y)\varphi(y)\d y$ with $K\in\Cc(\R^d\times\R^d;\C)$, then
\begin{equation*}
\Vert A\Vert_{L^2\to L^2}\leq\left(\sup_{x\in\R^d}\int_{\R^d}|K(x,y)|\d y\right)^{1/2}\left(\sup_{y\in\R^d}\int_{\R^d}|K(x,y)|\d x\right)^{1/2}.
\end{equation*}
\end{lem}

First of all, we have this rather crude result.

\begin{lem} \label{crudenorm}
Let $u\in S[(0,m^q,m^p);(2d,d,d)]$ with $\max(m^q,m^p)<-d$, then $\Ii^\h(\kappa;u;\Theta^x,\Theta^y)$ can be extended in a unique way to a linear bounded operator $L^2(\R^d;\C)\to L^2(\R^d;\C)$. More precisely, there exists $C(m^q,m^p)>0$ such that
\begin{equation} \label{eq:crude1}
\Vert \Ii^\h(\kappa;u;\Theta^x,\Theta^y)\Vert_{L^2\to L^2}\leq C\frac{M[u;m^q,m^p]}{\h^d(\det\Re\Theta^x\det\Re\Theta^y)^{1/4}}
\end{equation}
where
\begin{equation*}
M[u;m^q,m^p]:=\sup_{(x,y,q,p)\in\R^{4d}}\left|<q>^{-m^q}<p>^{-m^p}u(x,y,q,p)\right|.
\end{equation*}
Moreover, if the support of $u$ is contained in $\R^{2d}\times B((q_0,p_0),r)$, there exists $C'>0$ such that
\begin{equation} \label{eq:crude2}
\Vert \Ii^\h(\kappa;u;\Theta^x,\Theta^y)\Vert_{L^2\to L^2}\leq C'\frac{r^{2d}\Vert u\Vert_{L^\infty}}{\h^d(\det\Re\Theta^x\det\Re\Theta^y)^{1/4}}.
\end{equation}
\end{lem}

\begin{rem}
$ $
\begin{enumerate}[{\rm(i)}]
\item
\eqref{eq:crude2} is scale invariant but \eqref{eq:crude1} is not: the weaker scale invariant result would be
\begin{equation*}
\Vert \Ii^\h(\kappa;u;\Theta^x,\Theta^y)\Vert_{L^2\to L^2}\leq C\frac{M[u;m^q,m^p]}{\h^{-(m^q+m^p)/2}(\det\Re\Theta^x\det\Re\Theta^y)^{1/4}}.
\end{equation*}
\item
A crude estimate analogous to \eqref{eq:crude2} is used in~{\rm\cite{[LaptevSigal]}} to get estimate {\rm(2.14)} of Theorem 2.1.
\end{enumerate}
\end{rem}

\begin{dem}
By very crude estimates, we have
\begin{eqnarray*}
\lefteqn{\sup_{x\in\R^d}\int_{\R^d}|K^\h(\kappa;u;\Theta^x,\Theta^y)(x,y)|\d y\leq\frac{M[u;m^q,m^p]}{(2\pi\h)^{3d/2}}} \\ 
 & & \hspace{-0.5cm} \times\int_{\R^{2d}}\frac{\left(\int_{\R^d}e^{-\frac{1}{2\h}|(\Re\Theta^y)^{1/2}(y-q)|^2}\d y\right)\left(\sup_{x\in\R^d}e^{-\frac{1}{2\h}|(\Re\Theta^x)^{1/2}(x-X^\kappa(q,p))|^2}\right)}{<q>^{-m^q}<p>^{-m^p}}\d q\d p
\end{eqnarray*}
and the same holds true exchanging the role of $x$ and $y$ in $\sup_{x\in\R^d}$ and $\int_{\R^d}\d y$. Hence the required continuity and \eqref{eq:crude1} by application of Schur's Lemma. \eqref{eq:crude2} is proven the same way by substituting $\Vert u\Vert_{L^\infty}$ for $M[u;m^q,m^p]$ and the indicator function of $B((q_0,p_0),r)$ for $<q>^{m^q}<p>^{m^p}$ in the preceding estimate.
\end{dem}

As in the proof of the Calder\'on-Vaillancourt Theorem we will use the following Cotlar-Stein Lemma (for a proof see~\cite{[Martinez]} pp.48-49).

\begin{lem}[Cotlar-Stein]
Let $\Hh$ be a Hilbert space, $\omega:\Z^d\to\R$ and $(\Ii_\Gamma)_{\Gamma\in\Z^d}$ a family of bounded operators on $\Hh$ satisfying
\begin{equation*}
\forall\Gamma,\Gamma'\in\Z^d, \qquad \Vert\Ii_\Gamma^*\Ii_{\Gamma'}\Vert+\Vert\Ii_\Gamma\Ii_{\Gamma'}^*\Vert\leq\omega(\Gamma-\Gamma')
\end{equation*}
and
\begin{equation*}
\sum_{\Gamma\in\Z^d}\sqrt{\omega(\Gamma)}<\infty.
\end{equation*}
Then the series $\sum_{\Gamma\in\Z^d}\Ii_\Gamma$ is strongly convergent to a bounded operator $\Ii_\infty$ such that
\begin{equation*}
\Vert\Ii_\infty\Vert\leq\sum_{\Gamma\in\Z^d}\sqrt{\omega(\Gamma)}.
\end{equation*}
\end{lem}

To shorten notation, we introduce the following quantity
\begin{eqnarray*}
\Nn^\h(\kappa;u,v;\Theta^x,\Theta^y) & = & \Vert\Ii^\h(\kappa;v;\Theta^x,\Theta^y)^*\Ii^\h(\kappa;u;\Theta^x,\Theta^y)\Vert_{L^2\to L^2} \\
 & & +\Vert\Ii^\h(\kappa;v;\Theta^x,\Theta^y)\Ii^\h(\kappa;u;\Theta^x,\Theta^y)^*\Vert_{L^2\to L^2}
\end{eqnarray*}
where $u$ and $v$ are symbols of class $S[(0,-\infty);(2d,2d)]$ compactly supported in $(q,p)$ and first focus on the situation of symbols independent of $(x,y)$.

\begin{prop} \label{prop:compact_symbol_bound}
Let $u,v\in S[-\infty;2d]$ be two compactly supported symbols independent of $(x,y)$ with support denoted by $K_u$ and $K_v$ respectively. Then, $\Ii^\h(\kappa;u;\Theta^x,\Theta^y)$ has an $\h$-independent $L^2$ norm bound, more precisely,
\begin{enumerate}[{\rm(i)}]
\item
there exists $C>0$ such that
\begin{equation} \label{eq:compact1}
\hspace{-1.05cm}\Vert\Ii^\h(\kappa;v;\Theta^x,\Theta^y)^*\Ii^\h(\kappa;u;\Theta^x,\Theta^y)\Vert_{L^2\to L^2}\leq C\frac{e^{-\frac{\delta_{\Lambda(\Theta^x)\kappa}[K_u,K_v]^2}{4\h}}\Vert u\Vert_{L^1}\Vert v\Vert_{L^1}}{\h^{2d}(\det\Re\Theta^x\det\Re\Theta^y)^{1/2}}
\end{equation}
\begin{equation} \label{eq:compact1bis}
\hspace{-.9cm}\Vert\Ii^\h(\kappa;v;\Theta^x,\Theta^y)\Ii^\h(\kappa;u;\Theta^x,\Theta^y)^*\Vert_{L^2\to L^2}\leq C\frac{e^{-\frac{\delta_{\Lambda(\overline{\Theta^y})}[K_u,K_v]^2}{4\h}}\Vert u\Vert_{L^1}\Vert v\Vert_{L^1}}{\h^{2d}(\det\Re\Theta^x\det\Re\Theta^y)^{1/2}}
\end{equation}
where
\begin{equation} \label{eq:delta}
\delta_{\kappa'}[K_u,K_v]=\inf_{a\in\kappa'(K_u),b\in\kappa'(K_v)}|a-b|
\end{equation}
is the Hausdorff distance between $\kappa'(K_u)$ and $\kappa'(K_v)$,
\item
if $(K_u\cup K_v)\subset B((q_0,p_0),r)$ for $r\leq r_\infty$ with some fixed $r_\infty$, there exists $C'(\kappa,\Theta^x,\Theta^y,r_\infty)>0$ depending only on $M^\kappa_k$ for $k\leq d+2$, $\Vert\Lambda(\Theta^x)\Vert$, $\Vert\Lambda(\Theta^y)\Vert$, $\Vert\Lambda(\Theta^x)^{-1}\Vert$, $\Vert\Lambda(\Theta^y)^{-1}\Vert$ and $r_\infty$ such that
\begin{equation} \label{eq:compact2}
\Nn^\h(\kappa;u,v;\Theta^x,\Theta^y)\leq C'r^d\frac{\Vert u\Vert_{W^{d+1,\infty}}\Vert v\Vert_{W^{d+1,\infty}}}{(\det\Re\Theta^x\det\Re\Theta^y)^{1/2}}
\end{equation}
where
\begin{equation*}
\Vert w\Vert_{W^{d+1,\infty}}:=\sum_{|\alpha|\leq d+1}\Vert\partial_{(q,p)}^\alpha w\Vert_{L^\infty}.
\end{equation*}
\end{enumerate}
\end{prop}

\begin{rem}
$ $
\begin{enumerate}[{\rm(i)}]
\item
\eqref{eq:compact1} and \eqref{eq:compact1bis} are scale invariant but \eqref{eq:compact2} is not because of the factor $r^d$.
\item
An estimate analogous to \eqref{eq:compact2} is used in the proof of Theorem~7 in~{\rm\cite{[Butler]}} without explicit justification (a simple adaptation of the proof presented here provides one).
\end{enumerate}
\end{rem}

\begin{dem}
As (i) is scale invariant, we will restrict to $\h=1$. By a straightforward computation the operator $(2\pi)^{3d}\Ii^1(\kappa;v;\Theta^x,\Theta^y)^*\Ii^1(\kappa;u;\Theta^x,\Theta^y)$ has kernel
\begin{equation*}
N(x,y)=\int_{\R^{5d}}e^{i\omega^\kappa(x,y,z,q_1,q_2,p_1,p_2)}u(q_1,p_1)\overline{v(q_2,p_2)}\d q_1\d q_2\d p_1\d p_2\d z
\end{equation*}
where
\begin{equation*}
\omega^\kappa(x,y,z,q_1,q_2,p_1,p_2)=\Phi^\kappa(z,y,q_1,p_1;\Theta^x,\Theta^y)-\overline{\Phi^\kappa(z,x,q_2,p_2;\Theta^x,\Theta^y)}.
\end{equation*}
Reorganizing terms and writing $[\psi]^a_b$ for $\psi(a)-\psi(b)$, $\omega^\kappa$ splits into
\begin{eqnarray*}
\omega^\kappa_1 & = & [S^\kappa-\Xi^\kappa\cdot X^\kappa]^{(q_1,p_1)}_{(q_2,p_2)}+[\Xi^\kappa]^{(q_1,p_1)}_{(q_2,p_2)}\cdot\frac{X^\kappa(q_1,p_1)+X^\kappa(q_2,p_2)}{2} \\
 & & -p_1\cdot(y-q_1)+p_2\cdot(x-q_2) \\
 & & +\frac{i}{2}\Big[(y-q_1)\cdot\Theta^y(y-q_1)+(x-q_2)\cdot\overline{\Theta^y}(x-q_2)\Big]
\end{eqnarray*}
\begin{eqnarray*}
\omega^\kappa_2 & = & \frac{i}{4}[X^\kappa]^{(q_1,p_1)}_{(q_2,p_2)}\cdot\Re\Theta^x[X^\kappa]^{(q_1,p_1)}_{(q_2,p_2)}
\end{eqnarray*}
\begin{eqnarray*}
\omega^\kappa_3 & = & [\Xi^\kappa+\Im\Theta^xX^\kappa]^{(q_1,p_1)}_{(q_2,p_2)}\cdot\left(z-\frac{X^\kappa(q_1,p_1)+X^\kappa(q_2,p_2)}{2}\right) \\
 & & +i\left|(\Re\Theta^x)^{1/2}\left(z-\frac{X^\kappa(q_1,p_1)+X^\kappa(q_2,p_2)}{2}\right)\right|^2.
\end{eqnarray*}
The integral with respect to $z$ can then be performed (Fourier transform of a Gaussian) to get
\begin{equation*}
N(x,y)=\frac{\pi^{d/2}}{(\det\Re\Theta^x)^{1/2}}\int_{\R^{4d}}e^{i\omega^\kappa_0(x,y,q_1,q_2,p_1,p_2)}u(q_1,p_1)\overline{v(q_2,p_2)}\d q_1\d q_2\d p_1\d p_2
\end{equation*}
with $\omega^\kappa_0$ given by
\begin{equation*}
\omega^\kappa_1(x,y,q_1,q_2,p_1,p_2)+\frac{i}{4}\left|\Lambda(\Theta^x)[\kappa]^{(q_1,p_1)}_{(q_2,p_2)}\right|^2.
\end{equation*}

We have
\begin{equation} \label{eq:boundfrombelow}
\Im\omega^\kappa_0=\frac{1}{2}\Big[|(\Re\Theta^y)^{1/2}(y-q_1)|^2+|(\Re\Theta^y)^{1/2}(x-q_2)|^2\Big]+\frac{1}{4}\left|\Lambda(\Theta^x)[\kappa]^{(q_1,p_1)}_{(q_2,p_2)}\right|^2.
\end{equation}

Thus,
\begin{equation*}
\sup_{z_1\in\R^d}\int_{\R^d}|N(x,y)|\d z_2\leq\frac{\pi^{d/2}(2\pi)^{d/2}}{(\det\Re\Theta^x\det\Re\Theta^y)^{1/2}}\Vert u\Vert_{L^1}\Vert v\Vert_{L^1}e^{-\delta_{\Lambda(\Theta^x)\kappa}[K_u,K_v]^2/4}
\end{equation*}
where $(z_1,z_2)$ stands for $(x,y)$ or $(y,x)$. Hence by Schur's lemma,
\begin{equation*}
\Vert \Ii^1(\kappa;v;\Theta^x,\Theta^y)^*\Ii^1(\kappa;u;\Theta^x,\Theta^y)\Vert\leq\frac{\Vert u\Vert_{L^1}\Vert v\Vert_{L^1}e^{-\delta_{\Lambda(\Theta^x)\kappa}[K_u,K_v]^2/4}}{\pi^{2d}2^{5d/2}(\det\Re\Theta^x\det\Re\Theta^y)^{1/2}}.
\end{equation*}

Because of Corollary~\ref{II*eqI*I} and $\Vert w^\kappa\Vert_{L^1}=\Vert w\Vert_{L^1}$, we easily deduce that
\begin{equation*}
\Vert\Ii^1(\kappa;v;\Theta^x,\Theta^y)\Ii^1(\kappa;u;\Theta^x,\Theta^y)^*\Vert\leq\frac{\Vert u\Vert_{L^1}\Vert v\Vert_{L^1}e^{-\delta_{\Lambda(\overline{\Theta^y})}[K_u,K_v]^2/4}}{\pi^{2d}2^{5d/2}(\det\Re\Theta^x\det\Re\Theta^y)^{1/2}}.
\end{equation*}

Finally, we will prove (ii). Putting back $\h$ into the game and denoting by $N^\h(x,y)$ the kernel of $(2\pi\h)^{3d}\Ii^\h(\kappa;v;\Theta^x,\Theta^y)^*\Ii^\h(\kappa;u;\Theta^x,\Theta^y)$, we start from
\begin{equation*}
(\pi\h)^{-d/2}(\det\Re\Theta^x)^{1/2}N^\h(x,y)=N^\h_>(x,y)+N^\h_<(x,y)
\end{equation*}
where
\begin{equation*}
N^\h_\gtrless(x,y)=\int_{|\xi|^2+|\zeta|^2\gtrless\mu|x-y|}\int_{\R^{2d}}e^{\frac{i}{\h}\tilde{\omega}^\kappa_0}u(q,p)\overline{v(q+\xi,p+\zeta)}\d q\d p\d\xi\d\zeta
\end{equation*}
with $\tilde{\omega}^\kappa_0(x,y,q,p,\xi,\zeta):=\omega^\kappa_0(x,y,q,q+\xi,p,p+\zeta)$ and $\mu>0$ will be chosen later. On one hand, the bound from below
\begin{equation*}
\Im\tilde{\omega}^\kappa_0\geq\frac{1}{2}|(\Re\Theta^y)^{1/2}(y-q)|^2+\frac{1}{4}\left|\Lambda(\Theta^x)\left[\kappa\right]^{(q+\xi,p+\zeta)}_{(q,p)}\right|^2
\end{equation*}
implies that
\begin{eqnarray*}
\left|N^\h_>(x,y)\right| & \leq & \int_{|\xi|^2+|\zeta|^2\geq\mu|x-y|}e^{-\frac{1}{4\h}c^2_{\Lambda(\Theta^x)\kappa}[|\xi|^2+|\zeta|^2]} \\
 & & \left(\int_{\R^{2d}}e^{-\frac{1}{2\h}|(\Re\Theta^y)^{1/2}(y-q)|^2}\left|u(q,p)\overline{v(q+\xi,p+\zeta)}\right|\d q\d p\right)\d\xi\d\zeta \\
 & \leq & \left(\int_{|\xi|^2+|\zeta|^2\geq\mu|x-y|}\hspace{-2cm}e^{-\frac{1}{4\h}c^2_{\Lambda(\Theta^x)\kappa}[|\xi|^2+|\zeta|^2]}\d\xi\d\zeta\right)\frac{\h^{d/2}\Vert u\Vert_{L^\infty_qL^2_p}\Vert v\Vert_{L^\infty_qL^2_p}}{(\det\Re\Theta^y)^{1/2}}
\end{eqnarray*}
and, on the other hand, for $N^\h_<(x,y)$ we want to perform integration by parts with respect to $p$ to gain decay in $|x-y|/\h$. To do this, we establish some estimates for the derivatives of the phase $\tilde{\omega}^\kappa_0$. 
\begin{eqnarray}
\lefteqn{\nabla_p\Re\tilde{\omega}^\kappa_0} \nonumber \\
 & = & (x-y)-\xi+\frac{X^\kappa_p(q,p)+X^\kappa_p(q+\xi,p+\zeta)}{2}[\Xi^\kappa]^{(q,p)}_{(q+\xi,p+\zeta)} \nonumber \\
 & & -\frac{\Xi^\kappa_p(q,p)+\Xi^\kappa_p(q+\xi,p+\zeta)}{2}[X^\kappa]^{(q,p)}_{(q+\xi,p+\zeta)} \nonumber \\
 & = & (x-y) \nonumber \\
 & & -\int^1_0\frac{(1-\tau)^2}{2}\sum_{|\alpha|=3}\frac{1}{\alpha!}\left[X^\kappa_p(0)\partial^\alpha\Xi^\kappa(\tau)-\Xi^\kappa_p(0)\partial^\alpha X^\kappa(\tau)\right](\xi,\zeta)^\alpha\d\tau \nonumber \\
 & & \hspace{-0.2cm} +\frac{1}{2}\int^1_0(1-\tau)\sum_{\substack{|\alpha|=1 \\ |\beta|=2}}\frac{1}{\beta!}\left[\partial^\alpha X^\kappa_p(0)\partial^\beta\Xi^\kappa(\tau)-\partial^\alpha\Xi^\kappa_p(0)\partial^\beta X^\kappa(\tau)\right](\xi,\zeta)^{\alpha+\beta}\d\tau \nonumber \\
 & & -\frac{1}{2}\int^1_0(1-\tau)\sum_{|\alpha|=2}\frac{1}{\alpha!}\left[\partial^\alpha X^\kappa_p(\tau)[\Xi^\kappa]^0_1-\partial^\alpha\Xi^\kappa_p(\tau)[X^\kappa]^0_1\right](\xi,\zeta)^\alpha\d\tau, \label{eq:magic}
\end{eqnarray}
where $(\tau)$ stands for $(q+\tau\xi,p+\tau\zeta)$ and $[\psi]^0_1$ for $[\psi]^{(q,p)}_{(q+\xi,p+\zeta)}$. As a consequence, we have
\begin{equation*}
|\nabla_p\tilde{\omega}^\kappa_0|\geq|\nabla_p\Re\tilde{\omega}^\kappa_0|\geq|x-y|-K|(\xi,\zeta)|^3
\end{equation*}
and
\begin{equation*}
\left|\partial^\alpha_p\nabla_p\Re\tilde{\omega}^\kappa_0\right|\leq K_{|\alpha|+1}|(\xi,\zeta)|^3
\end{equation*}
for $|\alpha|\geq1$ where $K$ (respectively $K_k$) is polynomial in $M^\kappa_l$ for $l\leq2$ (respectively $l\leq k+1$). Finally we have
\begin{equation*}
\left|\partial^\alpha_p\Im\tilde{\omega}^\kappa_0\right|=\frac{1}{4}\left|\partial^\alpha_p\left|\Lambda(\Theta^x)[\kappa]^{(q,p)}_{(q+\xi,p+\zeta)}\right|^2\right|\leq K'_{|\alpha|}\Vert\Lambda(\Theta^x)\Vert^2(|\xi|^2+|\zeta|^2)
\end{equation*}
for $|\alpha|\geq2$ where $K'_k$ is polynomial in $M^\kappa_l$ for $l\leq k$.

Now, we take $\mu$ small enough ($4Kr\mu\leq1$) so that, if
\begin{equation} \label{localized}
(q,p)\in K_u, \quad (q+\xi,p+\zeta)\in K_v \quad {\rm and} \quad \mu|x-y|\geq|\xi|^2+|\zeta|^2,
\end{equation}
then
\begin{equation*}
|\nabla_p\tilde{\omega}^\kappa_0|\geq\frac{1}{2}|x-y|
\end{equation*}
and, for $|\alpha|\geq2$,
\begin{equation*}
\left|\partial^\alpha_p\tilde{\omega}^\kappa_0\right|\leq\mu\sqrt{K_{|\alpha|}^24r^2+{K'_{|\alpha|}}^2\Vert\Lambda(\Theta^x)\Vert^4}|x-y|.
\end{equation*}
Thus, if we have~\eqref{localized}, we can introduce the well-defined first order differential operator
\begin{equation*}
L_p=\frac{1}{1+\h^{-1}|\nabla_p\tilde{\omega}^\kappa_0|}\left[1-i\frac{\nabla_p\overline{\tilde{\omega}^\kappa_0}}{|\nabla_p\tilde{\omega}^\kappa_0|}\cdot\nabla_p\right]
\end{equation*}
which is such that $L_p(e^{i\tilde{\omega}^\kappa_0/\h})=e^{i\tilde{\omega}^\kappa_0/\h}$ and
\begin{equation*}
\left|(L_p^\dagger)^k w\right|\leq\frac{M^{(p)}_k\mu^k\left[1+r^2+\Vert\Lambda(\Theta^x)\Vert^4\right]^{k/2}}{\left(1+\frac{|x-y|}{2\h}\right)^k}\sum_{|\alpha|\leq k}\left|\partial_p^\alpha w\right|
\end{equation*}
where $M^{(p)}_k$ is polynomial in $K_l$ and $K'_l$ for $l\leq k+1$. Hence
\begin{eqnarray*}
\lefteqn{\left|N^\h_<(x,y)\right|} \\
 & = & \left|\int_{|\xi|^2+|\zeta|^2\leq\mu|x-y|}\int_{\R^{2d}}e^{\frac{i}{\h}\tilde{\omega}^\kappa_0}(L_p^\dagger)^{d+1}\left[u(q,p)\overline{v(q+\xi,p+\zeta)}\right]\d q\d p\d\xi\d\zeta\right| \\
 & \leq & \left(\int_{|\xi|^2+|\zeta|^2\leq\mu|x-y|}\hspace{-2cm}e^{-\frac{1}{4\h}c^2_{\Lambda(\Theta^x)\kappa}[|\xi|^2+|\zeta|^2]}\d\xi\d\zeta\right)2^{d+1}\sum_{|\alpha|+|\beta|\leq d+1}\Vert\partial^\alpha_pu\Vert_{L^\infty_qL^2_p}\Vert\partial^\beta_pv\Vert_{L^\infty_qL^2_p} \\
  & & \times\frac{\h^{d/2}M^{(p)}_{d+1}\mu^{d+1}\left[1+r^2+\Vert\Lambda(\Theta^x)\Vert^4\right]^{(d+1)/2}}{(\det\Re\Theta^y)^{1/2}\left(1+\frac{|x-y|}{2\h}\right)^{d+1}}
\end{eqnarray*}
and finally
\begin{eqnarray*}
\lefteqn{\sup_{z_1\in\R^d}\int_{\R^d}\left|N^\h(x,y)\right|\d z_2\leq\frac{\h^{3d}\pi^{d/2}}{(\det\Re\Theta^x\det\Re\Theta^y)^{1/2}}\Vert u\Vert_{L^\infty_qH^{d+1}_p}\Vert v\Vert_{L^\infty_qH^{d+1}_p}} \\
 & & \hspace{-0.4cm}\times\left[\int_{|\rho|^2\geq\mu|\sigma|}\hspace{-0.8cm}e^{-\frac{1}{4}c^2_{\Lambda(\Theta^x)\kappa}|\rho|^2}\d\rho\d\sigma+M[\kappa;\Theta^x;r]\int_{|\rho|^2\leq\mu|\sigma|}\frac{e^{-\frac{1}{4}c^2_{\Lambda(\Theta^x)\kappa}|\rho|^2}}{\left(1+\frac{|\sigma|}{2}\right)^{d+1}}\d\rho\d\sigma\right]
\end{eqnarray*}
where $(z_1,z_2)$ stands for $(x,y)$ or $(y,x)$ and
\begin{equation}
M[\kappa;\Theta^x;r]=2^{d+1}M^{(p)}_{d+1}\mu^{d+1}\left[1+r^2+\Vert\Lambda(\Theta^x)\Vert^4\right]^{(d+1)/2}.
\end{equation}

One concludes using Schur's Lemma and Corollary~\ref{II*eqI*I} to get the corresponding estimate for $\Ii^\h(\kappa;v;\Theta^x,\Theta^y)\Ii^\h(\kappa;u;\Theta^x,\Theta^y)^*$.
\end{dem}

\begin{rem}
In the proof of {\rm(ii)}, the use of Corollary \ref{II*eqI*I} is essential: if one tries to prove the second estimate directly, the identity analogous to~\eqref{eq:magic} is
\begin{equation*}
\nabla_p\Re\tilde{\omega}^\kappa_0=-\Xi^\kappa_p(q,p)(x-y)+O(|(\xi,\zeta)|^3)
\end{equation*}
which has a bad behaviour for non-invertible $\Xi^\kappa_p$.
\end{rem}

We extend this result to the situation of $(x,y)$-dependent symbols.

\begin{prop} \label{prop:xy_dep_compact_symbol_bound}
Let $u,v\in S[(0,-\infty);(2d,2d)]$ be two symbols supported in a compact set in $(q,p)$ independently of $(x,y)$, more precisely we will assume that the support of $u$ (resp. $v$) is contained in $\R^{2d}\times K_u$ (resp. $\R^{2d}\times K_v$) where $K_u$ and $K_v$ are compact subsets of $\R^d\times\R^d$. Then, for any $l\geq0$, there exists $C_l>0$ such that
\begin{eqnarray}
\lefteqn{\Vert\Ii^\h(\kappa;v;\Theta^x,\Theta^y)^*\Ii^\h(\kappa;u;\Theta^x,\Theta^y)\Vert\leq C_l} \nonumber \\
 & & \hspace{2cm}\times\frac{{\displaystyle\sum_{|\alpha|+|\beta|\leq l}}(\lambda^x)^{-\frac{|\alpha|+|\beta|}{2}}\Vert\partial^\alpha_xu\Vert_{L^\infty_{(x,y)}L^1_{(q,p)}}\Vert\partial^\beta_xv\Vert_{L^\infty_{(x,y)}L^1_{(q,p)}}}{\h^{2d}(\det\Re\Theta^x\det\Re\Theta^y)^{1/2}\left(1+\frac{\delta_{\Lambda(\Theta^x)\kappa}[K_u,K_v]^2}{\h}\right)^{l/2}} \label{eq:fullcrosssupp}
\end{eqnarray}
\begin{eqnarray}
\lefteqn{\Vert\Ii^\h(\kappa;v;\Theta^x,\Theta^y)\Ii^\h(\kappa;u;\Theta^x,\Theta^y)^*\Vert\leq C_l} \nonumber \\
 & & \hspace{2cm}\times\frac{{\displaystyle\sum_{|\alpha|+|\beta|\leq l}}(\lambda^y)^{-\frac{|\alpha|+|\beta|}{2}}\Vert\partial^\alpha_yu\Vert_{L^\infty_{(x,y)}L^1_{(q,p)}}\Vert\partial^\beta_yv\Vert_{L^\infty_{(x,y)}L^1_{(q,p)}}}{\h^{2d}(\det\Re\Theta^x\det\Re\Theta^y)^{1/2}\left(1+\frac{\delta_{\Lambda(\overline{\Theta^y})}[K_u,K_v]^2}{\h}\right)^{l/2}} \label{eq:fullcrosssuppbis}
\end{eqnarray}
where $\delta_{\kappa'}[K_u,K_v]$ is defined by \eqref{eq:delta} and
\begin{equation*}
\Vert w\Vert_{L^\infty_{(x,y)}L^1_{(q,p)}}:=\sup_{(x,y)\in\R^{2d}}\int_{\R^{2d}}|w(x,y,q,p)|\d q\d p.
\end{equation*}
\end{prop}

\begin{rem}
This result is scale invariant.
\end{rem}

\begin{dem}
We adapt the proof of Proposition~\ref{prop:compact_symbol_bound}: the main difference is that instead of computing explicitly the $z$-integral, we treat it like the Fourier transform of a Schwartz function and use that it has polynomial decay. The operator $(2\pi)^{3d}\Ii^1(\kappa;v;\Theta^x,\Theta^y)^*\Ii^1(\kappa;u;\Theta^x,\Theta^y)$ has kernel
\begin{equation*}
N(x,y)=\int_{\R^{5d}}e^{i\omega^\kappa(x,y,z,q_1,q_2,p_1,p_2)}u(z,y,q_1,p_1)\overline{v(z,x,q_2,p_2)}\d q_1\d q_2\d p_1\d p_2\d z
\end{equation*}
with
\begin{eqnarray*}
\hspace{-0.1cm}\Im\omega^\kappa & = & \frac{1}{2}|(\Re\Theta^y)^{1/2}(y-q_1)|^2+\frac{1}{4}\left|(\Re\Theta^x)^{1/2}[X^\kappa]^{(q_1,p_1)}_{(q_2,p_2)}\right|^2 \\
 & & \hspace{-0.2cm}+\frac{1}{2}|(\Re\Theta^y)^{1/2}(x-q_2)|^2+\left|(\Re\Theta^x)^{1/2}\left(z-\frac{X^\kappa(q_1,p_1)+X^\kappa(q_2,p_2)}{2}\right)\right|^2
\end{eqnarray*}
and
\begin{equation*}
\nabla_z\omega^\kappa=[\Xi^\kappa+\Im\Theta^xX^\kappa]^{(q_1,p_1)}_{(q_2,p_2)}+2i\Re\Theta^x\left(z-\frac{X^\kappa(q_1,p_1)+X^\kappa(q_2,p_2)}{2}\right).
\end{equation*}

We introduce the first order differential operator
\begin{equation*}
L_z=\frac{1}{1+|(\Re\Theta^x)^{-1/2}\nabla_z\omega^\kappa|^2}\left[1-i(\Re\Theta^x)^{-1}\overline{\nabla_z\omega^\kappa}\cdot\nabla_z\right]
\end{equation*}
which is such that $L_z(e^{i\omega^\kappa})=e^{i\omega^\kappa}$ and
\begin{equation*}
\left|(L_z^\dagger)^l w\right|\leq\frac{M^{(z)}_l}{\left(1+|(\Re\Theta^x)^{-1/2}\nabla_z\omega^\kappa|^2\right)^{l/2}}\sum_{|\alpha|\leq l}\left|[(\Re\Theta^x)^{-1/2}\nabla_z]^\alpha w\right|
\end{equation*}
where $M^{(z)}_l$ is a constant depending only on $l$. Hence the result, because of Schur's Lemma applied to
\begin{eqnarray*}
\lefteqn{\sup_{x\in\R^d}\int_{\R^d}|N(x,y)|\d y} \\
 & = & \sup_{x\in\R^d}\int_{\R^d}\left|\int_{\R^{5d}}e^{i\omega^\kappa}(L_z^\dagger)^l\left[u(z,y,q_1,p_1)\overline{v(z,x,q_2,p_2)}\right]\d q_1\d q_2\d p_1\d p_2\d z\right|\d y \\
 & \leq & M^{(z)}_l\tilde{M}_l2^{ld}\frac{{\displaystyle\sum_{|\alpha|+|\beta|\leq l}}(\lambda^x)^{-\frac{|\alpha|+|\beta|}{2}}\Vert\partial^\alpha_xu\Vert_{L^\infty_{(x,y)}L^1_{(q,p)}}\Vert\partial^\beta_xv\Vert_{L^\infty_{(x,y)}L^1_{(q,p)}}}{\left(1+\delta_{\Lambda(\Theta^x)\kappa}[K_u,K_v]^2\right)^{l/2}} \\
 & & \hspace{-0.2cm}\times\left(\sup_{x\in\R^d}e^{-\frac{1}{2}|(\Re\Theta^y)^{1/2}x|^2}\right)\left(\int_{\R^d}e^{-\frac{1}{2}|(\Re\Theta^y)^{1/2}y|^2}\d y\right)\left(\int_{\R^d}e^{-|(\Re\Theta^x)^{1/2}z|^2}\d z\right) \\
 & \leq & \frac{M^{(z)}_l\tilde{M}_l2^{ld}}{\pi^d2^{d/2}}\frac{{\displaystyle\sum_{|\alpha|+|\beta|\leq l}}(\lambda^x)^{-\frac{|\alpha|+|\beta|}{2}}\Vert\partial^\alpha_xu\Vert_{L^\infty_{(x,y)}L^1_{(q,p)}}\Vert\partial^\beta_xv\Vert_{L^\infty_{(x,y)}L^1_{(q,p)}}}{(\det\Re\Theta^x\det\Re\Theta^y)^{1/2}\left(1+\delta_{\Lambda(\Theta^x)\kappa}[K_u,K_v]^2\right)^{l/2}}
\end{eqnarray*}
and the corresponding estimate exchanging the role of $x$ and $y$ in $\sup_{x\in\R^d}$ and $\int_{\R^d}\d y$ (where $\tilde{M}_l=\sup_{r\geq0}[e^{-r/4}(1+r)^{l/2}]$).
\end{dem}

Finally we prove our main $L^2$-boundedness result.

{\bf Proof of Theorem~\ref{theo:L2bound}:}
Because of the scale invariance explained in Remark~\ref{rmk:scaleinv}, it is enough to consider the case $\h=1$. We introduce a function $\chi\in\Cc^\infty_0(\R^{2d};[0,1]))$ such that
\begin{equation*}
\supp\chi\subset\left[-\frac{1}{2}-\nu,\frac{1}{2}+\nu\right]^{2d}=:K, \qquad \chi(z)=1 \quad {\rm for} \quad z\in\left[-\frac{1}{2}+\nu,\frac{1}{2}-\nu\right]^{2d}
\end{equation*}
\begin{equation*}
\rm{and} \qquad \sum_{\Gamma\in\Z^{2d}}\chi_\Gamma=1
\end{equation*}
where $\chi_\Gamma(z)=\chi(z-\Gamma)$ and $\nu\leq1/4$.
This gives rise to a family of bounded FIOs whose symbols are defined by
\begin{equation*}
u_\Gamma(x,y,q,p):=\chi_\Gamma(\kappa_0(q,p))u(x,y,q,p)
\end{equation*}
with $\kappa_0$ a canonical transformation of class $\Bb$ to be determined later.

By Proposition~\ref{prop:xy_dep_compact_symbol_bound}, we have
\begin{itemize}
\item
if $|\Gamma-\Gamma'|_\infty>1$ then
\begin{equation} \label{eq:omega1}
\hspace{-0.2cm}\Nn^1(\kappa;u_\Gamma,u_{\Gamma'};\Theta^x,\Theta^y)\leq\frac{\min(1,\lambda^x,\lambda^y)^{-l/2}}{\left(1+\delta_{\Gamma,\Gamma'}^2\right)^{l/2}}\frac{C_l(1+2\nu)^{4d}\Vert u\Vert_{W^{l,\infty}_{(x,y)}L^\infty_{(q,p)}}^2}{(\det\Re\Theta^x\det\Re\Theta^y)^{1/2}}
\end{equation}
where
\begin{equation*}
\delta_{\Gamma,\Gamma'}=\min\left(\delta_{\Lambda(\Theta^x)\kappa\kappa_0^{-1}}[\supp\chi_\Gamma,\supp\chi_{\Gamma'}],\delta_{\Lambda(\overline{\Theta^y})\kappa_0^{-1}}[\supp\chi_\Gamma,\supp\chi_{\Gamma'}]\right),
\end{equation*}
\item
if $|\Gamma-\Gamma'|_\infty\leq1$ then
\begin{equation*}
(\supp\chi_\Gamma\cup\supp\chi_{\Gamma'})\subset B\left(\frac{\Gamma+\Gamma'}{2},\sqrt{2d}(1+2\nu)\right)
\end{equation*}
and
\begin{equation} \label{eq:omega2}
\Nn^1(\kappa;u_\Gamma,u_{\Gamma'};\Theta^x,\Theta^y)\leq C_0(2d)^{2d}(1+2\nu)^{4d}\frac{\Vert u\Vert_{L^\infty}^2}{(\det\Re\Theta^x\det\Re\Theta^y)^{1/2}}.
\end{equation}
\end{itemize}

Let $\omega(\Gamma-\Gamma')$ denote the right-hand side of \eqref{eq:omega1} and \eqref{eq:omega2}, as the number of $\Gamma$ such that $|\Gamma|_\infty\leq k$ is $(2k+1)^{2d}$, we have
\begin{eqnarray*}
\lefteqn{\frac{(\det\Re\Theta^x\det\Re\Theta^y)^{1/4}}{(1+2\nu)^{2d}\Vert u\Vert_{W^{l,\infty}_{(x,y)}L^\infty_{(q,p)}}}\sum_{\Gamma\in\Z^{2d}}\sqrt{\omega(\Gamma)}} \\
 & \leq & (18d)^d\sqrt{C_0}+\frac{\sqrt{C_l}}{\min(1,\lambda^x,\lambda^y)^{l/4}}\sum_{k\geq2}\frac{(2k+1)^{2d}-(2k-1)^{2d}}{\left(1+\eta_{[\kappa,\Theta^x,\Theta^y]}^2[k-(1+2\nu)]^2\right)^{l/4}} \\
 & \leq & 18^d\left[d^d\sqrt{C_0}+2^{3d-1}\frac{\sqrt{C_l}}{\min(1,\lambda^x,\lambda^y)^{l/4}}\sum_{k\geq1}\frac{k^{2d-1}}{\left(1+\eta_{[\kappa,\Theta^x,\Theta^y]}^2\frac{k^2}{2}\right)^{l/4}}\right]
\end{eqnarray*}
where
\begin{equation} \label{eq:eta}
\eta_{[\kappa,\Theta^x,\Theta^y]}=\inf_{\kappa_0}\left(c_{\Lambda(\Theta^x)\kappa\kappa_0^{-1}},c_{\Lambda(\overline{\Theta^y})\kappa_0^{-1}}\right).
\end{equation}
Hence the result on $\Ii^1(\kappa;u;\Theta^x,\Theta^y)$ by the Cotlar-Stein Lemma (the choice of $l=4d+1$ being the smallest that implies convergence of the series in~$k$).
\hspace {\stretch{1} }$\Box$

%%%%%%%%%%%%%%%%%%%%%%%%%%%%%%%%%%%%%%%%%%%%%%%%%%%%%%%%%%%%%%%%%%%%%%%%%%%%%%%%%%%%%%%%%
\section*{Appendix: Gaussian Integrals and Square Root of Matrices}
%%%%%%%%%%%%%%%%%%%%%%%%%%%%%%%%%%%%%%%%%%%%%%%%%%%%%%%%%%%%%%%%%%%%%%%%%%%%%%%%%%%%%%%%%

We consider the convex cone $\Cc$ of complex symmetric matrices with positive definite real part. Every matrix of $\Cc$ is invertible and has its spectrum included in the open half plane $\{z|\Re z>0\}$. It follows from matrix theory (see~\cite{[JohnsonOkuboReams]}) that each matrix of $\Cc$ admits an unique square root in $\Cc$. Furthermore, the square root of $M$ is given by the Dunford-Taylor integral (see~\cite{[Kato]} I.\textsection 5.6)
\begin{equation*}
M^{1/2}=\frac{1}{2i\pi}\int_\Gamma z^{1/2}(M-z)^{-1}\d z
\end{equation*}
where the integration path is a closed contour in the half-plane $\{z|\Re z>0\}$ making a turn around each eigenvalue in the positive direction and the value of $z^{1/2}$ is chosen so that it is positive for real positive $z$. As a consequence, the square root $M^{1/2}$ is an holomorphic function of $M$. Then, if one considers the computation of the Gaussian integral
\begin{equation*}
\frac{1}{(2\pi\h)^{d/2}}\int_{\R^d}e^{-\frac{M}{2\h}x\cdot x}\d x,
\end{equation*}
it is well known that, for positive definite real symmetric $M$, its value is given by $(\det M)^{1/2}=\det(M^{1/2})$. It directly follows from above that this property extends to any matrix $M\in\Cc$ (see Appendix A in~\cite{[Folland]} or Section 3.4. in~\cite{[Hormander3]} for an alternative explanation).

%%%%%%%%%%%%%%%%%%%%%%%%%%%%%%%%%%%%%%%%%%%%%%%%%%%%%%%%%%%%%%%%%%%%%%%%%%%%%%%%%%%%%%%%%
% References
%%%%%%%%%%%%%%%%%%%%%%%%%%%%%%%%%%%%%%%%%%%%%%%%%%%%%%%%%%%%%%%%%%%%%%%%%%%%%%%%%%%%%%%%%


\begin{thebibliography}{99}

\bibitem{[AsadaFujiwara]} K.~Asada and D.~Fujiwara, On some oscillatory integral transformation in $L^2(\R^n)$. {\it Jap. J. Math.}, {\bf 4} (2), pp.299-361 (1978).

\bibitem{[BilyRobert]} J.-M.~Bily and D.~Robert, The Semi-classical Van-Vleck Formula. Application to the Aharonov-Bohm Effect, Graffi, Sandro (ed.) et al., {\it Long time behaviour of classical and quantum systems}, Proceedings of the Bologna APTEX international conference, Bologna, Italy, September 13-17, 1999, Singapore: World Scientific, {\it Ser. Concr. Appl. Math.}, {\bf 1}, pp.89-106 (2001).

\bibitem{[Bony]} J.-M.~Bony, Op\'erateurs int\'egraux de Fourier et calcul de Weyl-H\"ormander (cas d'une m\'etrique symplectique), Journ\'ees EDP Saint-Jean-de-Monts, IX, pp.1-14 (1994).

\bibitem{[Bony1]} J.-M.~Bony, Evolution equations and microlocal analysis, in {\it Hyperbolic problems and related topics},
{\it Grad. Ser. Anal.}, International Press, Somerville, MA, pp.17-40 (2003).

\bibitem{[Butler]} J.~Butler, Global $h$ Fourier integral operators with complex-valued phase functions, {\it Bull. London Math. Soc.}, {\bf 34} (4), pp.479-489 (2002).

\bibitem{[CalderonVaillancourt]} A.P.~Calder\'on and R.~Vaillancourt, On the boundedness of pseudo-differential operators, {\it J. Math. Soc. Japan}, {\bf 23}, pp.374-378 (1971).

\bibitem{[CordobaFefferman]} A.~Cordoba and C.~Fefferman, Wave packets and Fourier integral operators, {\it Comm. in P.D.E.}, {\bf 3}, pp.979-1005 (1978).

\bibitem{[Coriasco]} S.~Coriasco, Fourier Integral Operators in SG Classes I: Composition Theorems and Action on SG Sobolev Spaces, {\it Rend. Sem. Mat. Univ. Politec. Torino}, {\bf 57} (4), pp.249-302 (1999).

\bibitem{[DimassiSjostrand]} M.~Dimassi and J.~Sj\"ostrand, Spectral Asymptotics in the Semi-Classical Limit, Lecture Notes Series 268, London Math. Society, Cambridge University Press (1999).

\bibitem{[DuistermaatHormander]} J.J.~Duistermaat and L.~H\"ormander, Fourier integral operators II, {\it Acta Math.}, {\bf 128} (3-4), pp.183-269 (1972).

\bibitem{[Folland]} G.B.~Folland, Harmonic Analysis in Phase Space, Annals of Mathematics Studies 122, Princeton University Press (1989).

\bibitem{[Fujiwara]} D.~Fujiwara, On the boundedness of integral transformations with highly oscillatory kernels, {\it Proc. Japan Acad.}, {\bf 51}, pp.96-99 (1975).

\bibitem{[Fujiwara1]} D.~Fujiwara, A construction of the fundamental solution for the Schr\"odinger equation, {\it J. d'Analyses Math.}, {\bf 35}, pp.41-96 (1979).

\bibitem{[Hormander]} L.~H\"ormander, Fourier integral operators I, {\it Acta Math.}, {\bf 127} (1-2), pp.79-183 (1971).

\bibitem{[Hormander1]} L.~H\"ormander, Oscillatory integrals and multipliers $FL^p$, {\it Ark. Mat.}, {\bf 11}, pp.1-11 (1973).

\bibitem{[Hormander2]} L.~H\"ormander, $L^2$ estimates for Fourier integral operators with complex phase, {\it Ark. Mat.}, {\bf 21}, pp.283-307 (1983).

\bibitem{[Hormander3]} L.~H\"ormander, The Analysis of Linear Partial Differential Operators I, Springer-Verlag, New York (1983).

\bibitem{[JohnsonOkuboReams]} C.R.~Johnson, Y.~Okubo, R.~Reams, Uniqueness of Matrix Square Roots and an Application, {\it Lin. Alg. Appl.}, {\bf 323}, pp.51-60 (2001).

\bibitem{[Kato]} T.~Kato, Perturbation Theory for Linear Operators, Springer-Verlag, New York (1966).

\bibitem{[Kay]} K.~Kay, Integral expressions for the semi-classical time-dependent propagator, {\it J. Chem. Phys.}, {\bf 100} (6), pp.4377-4392 (1994).

\bibitem{[KumanoGo]} H.~Kumano-Go, A calculus of Fourier integral operators on $\R^n$ and the fundamental solution for an operator of hyperbolic type, {\it Comm. P.D.E}, {\bf 1}, pp.1-44 (1976).

\bibitem{[LaptevSigal]} A.~Laptev and I.M.~Sigal, Global Fourier Integral Operators and semiclassical asymptotics, {\it Review of Math. Phys.}, {\bf 12} (5), pp.749-766 (2000).

\bibitem{[Lerner]} N.~Lerner, The Wick calculus of pseudo-differential operators and some of its applications, {\it Cubo Mat. Educ.}, {\bf 5}, pp.213-236 (2003).

\bibitem{[Martinez]} A.~Martinez, An Introduction to Semiclassical and Microlocal Analysis, Universitext, Springer-Verlag, New York (2002).

\bibitem{[Maslov]} V.P.~Maslov, Perturbation theory and asymptotic methods, (Russian) M.G.U., Moscow (1965).

\bibitem{[MelinSjostrand]} A.~Melin and J.~Sj\"ostrand, Fourier integral operators with complex valued phase functions, Springer Lecture Notes in Mathematics 459, pp.110-223 (1974).

\bibitem{[MelinSjostrand1]} A.~Melin and J.~Sj\"ostrand, Fourier integral operators with complex phase functions and parametrix for an interior boundary problem, {\it Comm. P.D.E.}, {\bf 1}, pp.313-400 (1976).

\bibitem{[Robert]} D.~Robert, Autour de l'approximation semi-classique, Progress in Mathematics 68, Birkh\"auser (1987).

\bibitem{[RuzhanskySugimoto]} M.~Ruzhansky and M.~Sugimoto, Global $L^2$-boundedness Theorems for a class of Fourier Integral Operators, {\it Comm. P.D.E.}, {\bf 31}, pp.547-569 (2006).

\bibitem{[SwartRousse]} T.~Swart and V.~Rousse, A Mathematical Justification for the Herman-Kluk Propagator, {\it preprint available on arXiv at} {\verb#http://arxiv.org/abs/0712.0752#}.

\bibitem{[Tataru]} D.~Tataru, Phase space transforms and microlocal analysis, in {\it Phase space analysis of partial differential equations} Vol.~II (Pubbl. Cent. Ric. Mat. Ennio Giorgi, Scuola Norm. Sup. Pisa), pp.505-524 (2004).

\bibitem{[Unterberger]} A.~Unterberger, Oscillateur harmonique et op\'erateurs pseudodiff\'erentiels, {\it Annales de l'Institut Fourier}, {\bf 29} (3), pp.201-221 (1979).

\bibitem{[Unterberger1]} A.~Unterberger, Les op\'erateurs m\'etadifferentiels, in {\it Complex analysis, microlocal calculus and relativistic quantum theory} (Proc. Internat. Colloq., Centre Phys., Les Houches, 1979), Lecture notes in Physics 126, pp.205-241 (1980).

\end{thebibliography}
\end{document}